\documentclass[12pt]{article}

\usepackage{latexsym,amscd}
\usepackage{bbm}
\usepackage{amsopn}
\usepackage{amsfonts}
\usepackage{amssymb}
\usepackage{amsthm}
\usepackage{amsmath}
\usepackage{psfrag}

\usepackage{epsfig,graphics}
\usepackage{t1enc}
\usepackage[cp1250]{inputenc}
\usepackage{a4wide}

\newtheorem{Theorem}{Theorem}[section]
\newtheorem{Proposition}[Theorem]{Proposition}
\newtheorem{Lemma}[Theorem]{Lemma}
\newtheorem{Corollary}[Theorem]{Corollary}

\theoremstyle{remark}

\newtheorem{Remark}{Remark}

\newcommand{\ddmatrix}[9]{\left(\begin{array}{ccc}   %3x3-Matrix
#1&#2&#3\\#4&#5&#6\\#7&#8&#9                %
\end{array}\right)}

\newcommand{\zzmatrix}[4]{\left(\begin{array}{cc}   %2x2-Matrix
#1&#2\\#3&#4\end{array}\right)}

\newcommand{\cfg}{{\mc X}}
\newcommand{\fdom}{{\mc A}}
\newcommand{\fdomll}{\mc A^{2}}
\newcommand{\fdoml}{\mc A^{1}}
\newcommand{\fdomo}{\mc A^{0}}
\newcommand{\fdomp}{\mc A^{p}}
\newcommand{\fdomq}{\mc A^{q}}
\newcommand{\fdompp}{\mc A^{p'}}
\newcommand{\fdomqp}{\mc A^{q'}}

\newcommand{\intfdom}{\dot{\mc A}{}}

\newcommand{\intfdomp}{\dot{\mc A}{}^{p}}
\newcommand{\intfdomq}{\dot{\mc A}{}^{q}}
\newcommand{\intC}{\dot C{}}
\newcommand{\TUT}{T\backslash \mr U(3) / T}
\newcommand{\UUT}{U_i\backslash \mr U(3) / T}
\newcommand{\UiUT}{U_i\backslash \mr U(3) / T}

\newcommand{\TUU}{T\backslash \mr U(3) / U_i}
\newcommand{\TUUi}{T\backslash \mr U(3) / U_i}
\newcommand{\TUUj}{T\backslash \mr U(3) / U_j}
\newcommand{\UUU}{U_i\backslash \mr U(3) / U_j}
\newcommand{\UiUUj}{U_i\backslash \mr U(3) / U_j}

\newcommand{\mTU}{\mu^{21}_{ij}}
\newcommand{\lTU}{\lambda^{21}_i}
\newcommand{\mUT}{\mu^{12}_{ij}}
\newcommand{\lUT}{\lambda^{12}_i}

\newcommand{\RR}{\mathbb{R}}

\newcommand{\CC}{\mathbb{C}}
\newcommand{\Z}{\mathbb{Z}}
\newcommand{\ZZ}{\mathbb{Z}}
\newcommand{\II}{\mathbbm{1}}
\newcommand{\diag}{{\rm diag}}
\newcommand{\mc}{\mathcal}

\newcommand{\mr}{\mathrm}

\newcommand{\vp}{\varphi}

\newcommand{\old}[1]{}

\newcommand{\GL}{{\mr{GL}}}
\newcommand{\gl}{{\mr{gl}}}

\newcommand{\SU}{{\mr{SU}}}

\newcommand{\U}{{\mr U}}

\newcommand{\rref}[1]{{\rm \ref{#1}}}
\newcommand{\ol}[1]{\overline{#1}}

\newcommand{\him}{{i\scriptscriptstyle -}}
\newcommand{\im}{{i_{_{\mbox{\footnotesize\rm -}}}}}
\newcommand{\hip}{{i\mbox{\tiny\rm +}}}
\newcommand{\ip}{{i_{_{\mbox{\!\tiny\rm +}}}}}

\newcommand{\imp}{{i_{_{\mp}}}}

\newcommand{\ipm}{{i_{_{\pm}}}}
\newcommand{\jm}{{j_{_{\mbox{\footnotesize\rm -}}}}}
\newcommand{\hjm}{{j\scriptscriptstyle -}}
\newcommand{\jp}{{j_{_{\mbox{\!\!\;\tiny\rm +}}}}}
\newcommand{\hjp}{{j\mbox{\tiny\rm +}}}
\newcommand{\jmp}{{j_{_{\mp}}}}

\newcommand{\jpm}{{j_{_{\pm}}}}

\newcommand{\km}{{k_{_{\mbox{\footnotesize\rm -}}}}}

\newcommand{\kp}{{k_{_{\mbox{\!\tiny\rm +}}}}}

\newcommand{\minus}{\mbox{\footnotesize -}}
\newcommand{\sign}{\mbox{\rm sign}}
\newcommand{\todo}[1]{}
\newcommand{\comment}[1]{}

\newcommand{\image}{\mathrm{im}\,}

\newcommand{\bre}{\begin{Remark}}
\newcommand{\ere}{\end{Remark}}

\newcommand{\bco}{\begin{Corollary}}
\newcommand{\eco}{\end{Corollary}}

\newcommand{\ble}{\begin{Lemma}}
\newcommand{\ele}{\end{Lemma}}

\newcommand{\bpr}{\begin{Proposition}}
\newcommand{\epr}{\end{Proposition}}

\newcommand{\btm}{\begin{Theorem}}
\newcommand{\etm}{\end{Theorem}}

\newcommand{\beq}{\begin{equation}}
\newcommand{\eeq}{\end{equation}}

\newcommand{\beqa}{\begin{eqnarray}}
\newcommand{\eeqa}{\end{eqnarray}}

\newcommand{\beqast}{\begin{eqnarray*}}
\newcommand{\eeqast}{\end{eqnarray*}}

\newcommand{\punkt}[1]{\put(#1){\circle*{0.075}}}
\newcommand{\linie}[3]{\put(#1){\line(#2){#3}}}
\newcommand{\vektor}[3]{\put(#1){\vector(#2){#3}}}
\newcommand{\marke}[3]{\put(#1){\put(0.05,0.1){\makebox(-0.1,-0.2)[#2]{$#3$}}}}

\begin{document}

\title{\bf On the Topology of %Cell Complex Structure for
the Reduced Classical Configuration
Space of Lattice QCD}

\author{
    S.~Charzy\'nski\\
    Center for Theoretical  Physics, Polish Academy of Sciences\\
    al. Lotnik\'ow 32/46, 02-668 Warsaw, Poland\\
    \ \\
    G.~Rudolph and M.~Schmidt\\
    Institut f\"ur Theoretische Physik, Universit\"at Leipzig\\
    Augustusplatz 10/11, 04109 Leipzig, Germany\\
    }

\maketitle

\vspace{2cm}

\\


\vspace{2cm}

\begin{abstract}

We study the topological structure of the quotient $\cfg$ of
$\SU(3)\times\SU(3)$ by diagonal conjugation. This is the simplest
nontrivial example for the classical reduced configuration space
of chromodynamics on a spatial lattice in the Hamiltonian
approach. We construct a cell complex structure of $\cfg$ in such
a way that the closures of strata are subcomplexes and
compute the homology and cohomology groups of the
strata and their closures.

\end{abstract}

\newpage

%\tableofcontents

%\newpage

%%%%%%%%%%%%%%%%%%%%%%%%%%%%%%%%%%%%%%%%%%%%%%%%%%%%%%%%%%%%%%%%%%%%%
%%%%%%%%%%%%%%%%%%%%%%%%%%%%%%%%%%%%%%%%%%%%%%%%%%%%%%%%%%%%%%%%%%%%%

\section{Introduction}
\label{Sintro}
\setcounter{equation}{0}

%%%%%%%%%%%%%%%%%%%%%%%%%%%%%%%%%%%%%%%%%%%%%%%%%%%%%%%%%%%%%%%%%%%%%
%%%%%%%%%%%%%%%%%%%%%%%%%%%%%%%%%%%%%%%%%%%%%%%%%%%%%%%%%%%%%%%%%%%%%

This paper is a continuation of \cite{confspace} where we have studied the
quotient of $\SU(3)\times\SU(3)$ by diagonal conjugation by means of invariant
theory. Let us start with some introductory remarks on the motivation for the
study of such a very specific space.

Many aspects of the rich mathematical and physical structure of
nonabelian gauge theories are not accessible by perturbation
theory, but only by nonperturbative methods. Perhaps the most
prominent of these aspects is the low energy hadron physics and,
in particular, quark confinement. Another such aspect is the
stratified structure of the classical configuration space of the
theory, i.e., of the space of orbits of the group of local gauge
transformations acting on the classical fields. This space
consists of an open dense manifold part, the 'principal stratum',
and several singularities which themselves decompose into
manifolds of varying dimension, the 'non-principal' or 'secondary
strata'. We note that similar structures arise in the study of the
geometry of quantum mechanical state spaces
\cite{adelman,Birkhoff}. While, by now, much is known about the
stratified structure itself, see \cite{RSV:ot,RSV:porev}, it is
still open whether and how it expresses itself in the physical
properties of the theory. A systematic investigation of this
question requires a concept of how to encode the stratification
into the quantum theory. A promising candidate is the concept of
costratified Hilbert space recently developed by Huebschmann
\cite{Hue:costrfHispa}. Very roughly, a costratified Hilbert space
consists of a total Hilbert space $\mc H$ and a family of Hilbert
spaces $\mc H_i$ associated with the closures of the strata,
together with a family of bounded linear maps $\mc H_i \to \mc
H_j$ whenever stratum $j$ is contained in the closure of stratum
$i$. As was demonstrated in \cite{Hue:costrfHispa}, one way to
construct such costratified Hilbert space is by extending methods
of geometric quantization to so-called stratified symplectic
spaces. (However, the concept of costratified Hilbert space does
not rely on any particular quantization procedure.) A stratified
symplectic space is a stratified space in the usual sense where,
in addition, the total space carries a Poisson structure, the
strata carry symplectic structures and the structures are
compatible in the sense that the injections of the strata into the
total space are Poisson maps. Such spaces naturally emerge as the
reduced phase spaces of Hamiltonian systems with symmetries by the
process of singular Marsden-Weinstein reduction
\cite{CushmanBates}. These observations suggest the following
strategy to construct a quantum gauge theory with the
stratification encoded \cite{howto}: formulate the classical
theory as a Hamiltonian system with symmetries, construct the
reduced phase space and then try to apply the method of stratified
K\"ahler quantization \comment{geometric quantization of
stratified
symplectic space}%
developed in \cite{Hue:costrfHispa} to obtain a costratified
Hilbert space. To separate the difficulties arising from infinite
dimensions from those related to symmetry reduction it is
reasonable to first work in the lattice approximation. For a simple model
motivated by $\SU(2)$-lattice gauge theory the program was carried out in
\cite{hrs}.

For the construction explained above, and of course for the
discussion of classical and quantum dynamics anyhow, it is
important to know the topological and geometric structure of the
reduced phase space as well as of the reduced configuration space.
In this work, we focus on the latter. Using a tree gauge, it can
be shown to be given by the quotient of several copies of the
(compact) gauge group by diagonal conjugation. Since we are
interested in QCD, our gauge group is $\SU(3)$. Since for one copy
of $\SU(3)$ the quotient is well known to be a Weyl alcove in the
subgroup of diagonal matrices, the simplest nontrivial case is $2$
copies. This corresponds to a lattice consisting of $2$
plaquettes. The description of the quotient of $2$ copies by means
of invariant theory was derived in a previous paper
\cite{confspace}. In the present work we construct a cell complex
structure which is compatible with the orbit type stratification and use it
to compute the homology and cohomology groups of the quotient
space as well as of the strata and their closures.

To conclude these introductory notes, let us remark the following.
The strategy explained above follows the path to first reduce the
symmetries and then quantize. One can as well follow the
alternative path of reducing the symmetries on the quantum level,
i.e., to start from a field algebra and to construct the algebra
of observables by implementing gauge invariance and the local
Gauss law, see \cite{KR:QCD}. To implement the stratification in
this approach, one has to work out the concept of (co)stratified
observable algebra and to modify the construction in \cite{KR:QCD}
so as to provide, in addition, observable algebras associated with
the strata. In a second step, one has to study representations of
the costratified observable algebra so obtained. Of course, such
representations act in a costratified Hilbert space. Also
for this step one should start from the results in \cite{KR:QCD}
and try to extend the methods used there to the costratified case.
Although this alternative strategy seems equally promising to us,
no work in this direction has been done yet.

The paper is organized as follows. In Section \rref{Sbasics} we
define the space we will study and introduce some notation. 
In Section \rref{Scfg} we relate the reduced configuration space to
certain double quotients of $\mr U(3)$ and show how to construct a
cell complex structure of the reduced configuration space from
cell decompositions of the double quotients. Section \rref{SY} is
devoted to the study of the most important of these double
quotients, i.e., $T\backslash\mr U(3)/T$, where $T$ denotes the
subgroup of $\mr U(3)$ of diagonal matrices. The remaining double
quotients and the factorization maps between them are studied in
Section \rref{Sother}. Based on these preparations, in Section
\rref{SCW} the desired cell complex structure of the reduced
configuration space is given and the boundary operator is
computed. In Section \rref{Sstrat} it is shown that the closures
of the strata are subcomplexes. Finally, in Section \rref{Shml}
the homology and cohomology groups of the reduced configuration
space and its strata are derived. We conclude with a brief
discussion.

%%%%%%%%%%%%%%%%%%%%%%%%%%%%%%%%%%%%%%%%%%%%%%%%%%%%%%%%%%%%%%%%%%%%%
%%%%%%%%%%%%%%%%%%%%%%%%%%%%%%%%%%%%%%%%%%%%%%%%%%%%%%%%%%%%%%%%%%%%%

\section{Basic notation}
\label{Sbasics}
\setcounter{equation}{0}

%%%%%%%%%%%%%%%%%%%%%%%%%%%%%%%%%%%%%%%%%%%%%%%%%%%%%%%%%%%%%%%%%%%%%
%%%%%%%%%%%%%%%%%%%%%%%%%%%%%%%%%%%%%%%%%%%%%%%%%%%%%%%%%%%%%%%%%%%%%

Let $\cfg$ denote the quotient space of the action of $\SU(3)$ on 
$\SU(3)\times\SU(3)$ by diagonal conjugation, 
$$
(a,b)\mapsto(gag^{-1},gbg^{-1})
 \,,~~~~~~
a,b,g\in\SU(3)\,.
$$
This space can be interpreted as the reduced configuration space of an
$\SU(3)$-lattice gauge model defined on a lattice consisting of two plaquettes.
See \cite{confspace} for details. The space $\cfg$ is stratified by the orbit types of
the action. The stratification will be explained in detail in Section
\rref{Sstrat}. The aim of this paper is to construct a cell decomposition of
$\cfg$ such that the closures of the strata are subcomplexes and to use it to
compute the homology and cohomology groups of the strata and their closures.

Let us introduce some notation. Consider the toral subgroup of
diagonal matrices $a$ in $\SU(3)$. The conditions $a_{11}=a_{22}$,
$a_{11}=a_{33}$, $a_{22}=a_{33}$ on the entries of $a$ define
one-parameter subgroups that cut this toral subgroup into 6 closed
triangular subsets. Let $\fdom$ denote one of them. The embedding
$\fdom\to\SU(3)$ descends to a homeomorphism of $\fdom$ onto the
quotient of $\SU(3)$ by inner automorphisms
(see \cite{confspace} for more details). $\fdom$ has a natural
cell decomposition consisting of a $2$-cell $\fdomll\equiv\fdom$,
three $1$-cells ('edges') $\fdoml_i$ and three $0$-cells
('vertices') $\fdomo_i$ which make up the center $\ZZ_3$ of
$\SU(3)$. We assume the $1$-cells to be oriented by the boundary
orientation induced from $\fdomll$. In the sequel, for the cells
of $\fdom$ we will loosely write $\fdom^p_i$, although for $p=2$
the index $i$ is redundant. The interior of $\fdom^p_i$ will be denoted by
$\intfdom^p_i$.

We will also work with $\mr U(3)$ and its subgroup of diagonal
matrices $T\cong \mr U(1)^3$, as well as the subgroups
$U_1,U_2,U_3\cong \mr U(1)\times\mr U(2)$, consisting of the
matrices
$$
\footnotesize
 \begin{pmatrix}
\alpha & 0 & 0 \\ 0 & b_{11} & b_{12} \\ 0 & b_{21} & b_{22}
 \end{pmatrix}
 \,,~~~~
 \begin{pmatrix}
b_{11} & 0 & b_{12} \\ 0 & \alpha & 0 \\ b_{21} & 0 & b_{22}
 \end{pmatrix}
 \,,~~~~
 \begin{pmatrix}
b_{11} & b_{12} & 0 \\ b_{21} & b_{22} & 0 \\ 0 & 0 & \alpha
 \end{pmatrix}
 \,,
$$
respectively, where $\alpha\in\mr U(1)$ and $ b\in\mr U(2)$.

Next, for $i=1,2,3$, let $\ipm := i\pm 1\mod 3$. The elements of
the permutation group $\mc S_3$ are denoted by $\tau$ or $\sigma$.
We number them in the following way: $\tau_0$ denotes the identity
permutation, $\tau_i$ for $i=1,2,3$ denotes the transposition of
$\ip$ and $\im$ and $\tau_4$ and $\tau_5$ are the backward and
forward cyclic permutations, respectively. A representation of
$\mc S_3$ in $\U(3)$ is given by the matrices
$$
\tau_{ij} = \delta_{\tau(i)\, j}
 \,,~~~~~~
\tau\in\mc S_3\,,
$$
where $\delta_{kl}$ denotes the Kronecker symbol. Explicitly,
$$
\footnotesize
 \begin{array}[b]{rclrclrcl}
\tau_0 & = & \ddmatrix{1}{0}{0}{0}{1}{0}{0}{0}{1}
 \,,~~ &
\tau_1 & = & \ddmatrix{1}{0}{0}{0}{0}{1}{0}{1}{0}
 \,,~~ &
\tau_2 & = & \ddmatrix{0}{0}{1}{0}{1}{0}{1}{0}{0}
\\[0.8cm]
\tau_3 & = & \ddmatrix{0}{1}{0}{1}{0}{0}{0}{0}{1}
 \,,~~ &
\tau_4 & = & \ddmatrix{0}{1}{0}{0}{0}{1}{1}{0}{0}
 \,,~~ &
\tau_5 & = & \ddmatrix{0}{0}{1}{1}{0}{0}{0}{1}{0}\,.
 \end{array}
$$
The centralizer and the normalizer of a subset $A\subseteq G$ will be denoted
by $\mr C_G(A)$ and $\mr N_G(A)$, respectively.

%%%%%%%%%%%%%%%%%%%%%%%%%%%%%%%%%%%%%%%%%%%%%%%%%%%%%%%%%%%%%%%%%%%%%
%%%%%%%%%%%%%%%%%%%%%%%%%%%%%%%%%%%%%%%%%%%%%%%%%%%%%%%%%%%%%%%%%%%%%

\section{The configuration space in terms of double quotients}
\label{Scfg}
\setcounter{equation}{0}

%%%%%%%%%%%%%%%%%%%%%%%%%%%%%%%%%%%%%%%%%%%%%%%%%%%%%%%%%%%%%%%%%%%%%
%%%%%%%%%%%%%%%%%%%%%%%%%%%%%%%%%%%%%%%%%%%%%%%%%%%%%%%%%%%%%%%%%%%%%

The starting point of our considerations is the map
$$
\vp :\fdom\times\fdom\times\mr U(3)\to \cfg
 ~,~~~~
(t,s,g) \mapsto [(t,gsg^{-1})]\,,
$$
where $[\,\cdot\,]$ denotes the class w.r.t.\ the diagonal action of $\SU(3)$.

\ble\label{Lvp}

The map $\vp$ is surjective and closed. Equality $\vp(t,s,g) =
\vp(t',s',g')$ holds iff $t=t'$, $s=s'$ and $g' = hgk$ for some
$h\in\mr C_{\mr U(3)} (t)$ and $k\in\mr C_{\mr U(3)}(s)$.

\ele

{\it Proof.}~ $\vp$ is surjective: Let
$(a,b)\in\SU(3)\times\SU(3)$. There exist $c,d\in\SU(3)$ such that
$t:=cac^{-1}$ and $s:=dbd^{-1}$ are in $\fdom$. Denote
$g:=cd^{-1}$. Then $(t,gsg^{-1}) = (cac^{-1}, cbc^{-1})$, hence
$\vp(t,s,g) = [(a,b)]$. 
$\vp$ is closed, because it is a map from a compact space to a
Hausdorff space. 
To determine the preimages, let $(t,s,g)$ and $(t',s',g')$ be given. If $t'$,
$s'$
and $g'$ are as in the lemma then they are obviously mapped by
$\vp$ to the same point in $\cfg$. Conversely, assume that
$\vp(t',s',g') = \vp(t,s,g)$. Then there exists $h\in\SU(3)$ such
that $t' = hth^{-1}$ and $g' s' g'^{-1} = h (gsg^{-1})h^{-1}$.
Since $t'$ and $t$ are both in $\fdom$, the first equality implies
$t' = t$, hence $h\in C_{\SU(3)}(t) \subseteq C_{\U(3)}(t)$.
Similarly, the second equality implies $g'^{-1}hg\in
C_{\U(3)}(s)$. Denoting $k := (g'^{-1}hg)^{-1}$ we obtain $g' =
hgk$, as asserted.
 \qed
\\

Thus, up to diagonal conjugacy, pairs $(a,b)\in\SU(3)\times\SU(3)$ can be
characterized by triples $(t,s,g)$ with $s$ and $t$ being uniquely determined
and $g$ representing a class in the double quotient
$
\mr C_{\mr U(3)}(t) \backslash \U(3) / \mr C_{\mr U(3)}(s)\,,
$
taken wrt.\ left and right multiplication, respectively. Since for any
$t$ in the interior of the cell $\fdom^p_i$ of $\fdom$, the centralizer is
$C_{\mr U(3)}(t) = C_{\mr U(3)}(\fdom^p_i)$, to any pair of cells $\fdom^p_i$,
$\fdom^q_j$ there corresponds a double quotient
$$
D(\fdom^p_i,\fdom^q_j) := \mr C_{\mr U(3)}(\fdom^p_i) \backslash
\U(3) / \mr C_{\mr U(3)}(\fdom^q_j)\,.
$$
In detail, we find
 \begin{align}\nonumber
D(\fdom^2,\fdom^2) & = \TUT
 \,,~~~
D(\fdom^1_i,\fdom^2) = \UUT
 \,,~~~
D(\fdom^2,\fdom^1_i) = \TUU
 \,,
\\ \nonumber
D(\fdom^1_i,\fdom^1_j) & = \UUU
 \,,
\\ \label{Gdoquos}
D(\fdom^0_i,\fdom^2) & = D(\fdom^2,\fdom^0_i) = D(\fdom^0_i,\fdom^1_j) =
D(\fdom^1_i,\fdom^0_j) = D(\fdom^0_i,\fdom^0_j) = \{\ast\}
 \,.
 \end{align}
There exist the following natural factorization maps:
 \beq
 \label{Gfactmaps}
\begin{CD}
\TUT
 @> \lambda^{12}_i >>
\UUT
\\
 @V \lambda^{21}_j VV @VV \mu^{12}_{ij} V
\\
\TUUj @> \mu^{21}_{ij} >> \UUU @>>> \{\ast\}
\end{CD}
 \eeq
 \comment{
 \beq\label{Gfactmaps}
\begin{array}{ccccccc}
& & \UUT & & & &
\\
& \overset{\lUT}{~}{\nearrow} & & {\searrow}\,\overset{\mUT}{~} & & &
\\
\TUT & & & & \UUU & \longrightarrow & \{\ast\}
\\
& \overset{\textstyle~}{\scriptstyle\lTU}\searrow & &
\nearrow\,\overset{~}{\scriptstyle\mTU} & & &
\\
& & \TUU & & & &
\end{array}
 \eeq
 }%
The idea of the construction of the cell decomposition of $\cfg$ can be stated
as follows.
According to Lemma \rref{Lvp}, $\vp$ induces maps
 \beq
 \label{Gdefpipqij}
\pi^{pq}_{ij} : \fdomp_i\times\fdomq_j\times D(\fdom^p_i,\fdom^q_j)
\to\cfg
 \eeq
which, since $\vp$ is closed, restrict to homeomorphisms of $\intfdom^p_i
\times \intfdom^q_j \times D(\fdom^p_i,\fdom^q_j)$ onto the image of this
subset in $\cfg$. Hence, if one has a cell $K$ in the double quotient
$D(\fdom^p_i,\fdom^q_j)$ then $\fdom^p_i\times\fdom^q_j\times K$ with
characteristic map obtained by restriction of $\pi^{pq}_{ij}$ is a candidate
for a cell of $\cfg$. The potential identifications on the boundary can be
kept track of
by exploring, for each boundary cell $\fdom^{p\minus 1}_{i'}$ of $\fdom^p_i$ and
$\fdom^{q\minus 1}_{j'}$ of $\fdom^q_j$, the relation between $\pi^{pq}_{ij}$
and the maps $\pi^{p\minus 1~q}_{i'j}$ and $\pi^{p~q\minus 1}_{ij'}$,
respectively. More generally, the situation is the following. Given
$\fdom^{p'}_{i'}$ and $\fdom^{q'}_{j'}$ such that
 \beq
 \label{GLTCW1}
\fdomp_i\subseteq\fdompp_{i'}
 \,,~~~~
\fdomq_j\subseteq\fdomqp_{j'}
 \,,
 \eeq
the factorization map $D(\fdom^{p'}_{i'},\fdom^{q'}_{j'})\to
D(\fdom^p_i,\fdom^q_j)$ exists--it is in fact a composition of some of the
maps \eqref{Gfactmaps}--and the diagram
 \beq
 \label{Gdgr}
\begin{CD}
\fdom^p_i\times\fdom^q_j \times D(\fdom^{p'}_{i'},\fdom^{q'}_{j'})
 @>>>
\fdom^{p'}_{i'}\times\fdom^{q'}_{j'} \times D(\fdom^{p'}_{i'},\fdom^{q'}_{j'})
\\
 @VVV @VV\pi^{p'q'}_{i'j'} V
\\
\fdomp_i\times\fdomq_j\times D(\fdom^p_i,\fdom^q_j) @>\pi^{pq}_{ij}>> \cfg
\end{CD}
 \eeq
commutes, where the upper horizontal arrow is given by the natural injection
and the left vertical arrow is given by the identical map of
$\fdom^p_i\times\fdom^q_j$ times the above factorization map. Thus, in effect
it is this factorization map which carries the information about the boundary
identifications carried out by the (prospective) characteristic map.

\btm\label{TCW}

Let there be given cell decompositions of the double quotients
\eqref{Gdoquos} such that the factorization maps \eqref{Gfactmaps} are
cellular. Then the collection of all products
$\fdom^p_i\times\fdom^q_j\times K$, where $\fdom^p_i$, $\fdom^q_j$
are cells of $\fdom$, $K$ is a cell of $D(\fdom^p_i,\fdom^q_j)$
and the characteristic map is defined by restriction of
$\pi^{pq}_{ij}$, defines a cell complex structure on $\cfg$. If
the cells are oriented by the natural product orientation, the
boundary operator is given by
 \begin{align*}
\partial(\fdom^p_i\times\fdom^q_j\times K)
 = &
\sum\nolimits_{i'} \fdom^{p\minus 1}_{i'} \times \fdom^q_j\times
\rho_{i'\,\ast}(K)
 +
(-1)^p\sum\nolimits_{j'} \fdom^p_i\times \fdom^{q\minus
1}_{j'}\times \sigma_{j'\,\ast}(K)
\\
 & +
(-1)^{p+q}\fdom^p_i\times\fdom^q_j\times \partial K
 \,,
 \end{align*}
where the sums run over the boundary cells of $\fdom^p_i$ and
$\fdom^q_j$, respectively, equipped with the correct sign, and $\rho_{i'}$ and
$\sigma_{j'}$ stand for the factorization
maps $D(\fdom^p_i,\fdom^q_j) \to D(\fdom^{p\minus 1}_{i'} , \fdom^q_j)$ and
$D(\fdom^p_i,\fdom^q_j) \to D(\fdom^p_i,\fdom^{q\minus 1}_{j'} )$, respectively, given
in \eqref{Gfactmaps}.

\etm

{\it Proof.}~ We give an inductive construction of the skeleta. For the sets
$\fdom^p_i\times\fdom^q_j\times K$ we will use the shorthand notation
$C^{pqr}_{ijk}$, where $r$ stands for the dimension of $K$ and
$k$ is a virtual label for the $r$-cells of $D(\fdom^p_i,\fdom^q_j)$. Moreover,
in order to distinguish between the boundary of $C^{pqr}_{ijk}$
in $\cfg$ and its boundary in the cell complex $\fdom^p_i\times\fdom^q_j\times
D(\fdom^p_i,\fdom^q_j)$, the first one will be denoted by $\partial$ and the
second one by $\tilde\partial$. In a sense, $\tilde\partial C^{pqr}_{ijk}$ is the
'natural' boundary of $C^{pqr}_{ijk}$.

We will say that a cell complex $\cfg^n$
has the property $(\ast)$ iff it is homeomorphic to the image of the
$n$-skeleton of $\fdom^2\times\fdom^2\times D(\fdom^2,\fdom^2)$
under $\pi^{22}$ (and can thus be identified with a subset of $\cfg$).

We start with defining $\cfg^0$ to consist of the nine isolated
points $C^{000}_{ij} = \fdomo_i\times\fdomo_j\times \{\ast\}$,
$i,j=1,2,3$. Then $\cfg^0$ is a cell complex and $(\ast)$
holds trivially.

Now assume that the $C^{pqr}_{ijk}$ with $p + q + r = n$
constitute a cell complex $\cfg^n$ which has the property $(\ast)$ and let
some $C^{pqr}_{ijk}$ with $p+q+r=n+1$ be given. Consider the diagram
\eqref{Gdgr} with $p' = q' = 2$.
Due to the assumption  that the factorization maps \eqref{Gfactmaps} are
cellular, the left vertical arrow in this diagram is cellular. Hence, the
preimage of $\tilde\partial C^{pqr}_{ijk}$ under this map is a union of
$n$-cells of $\fdom^2\times\fdom^2\times D(\fdom^2,\fdom^2)$ and is therefore
mapped by $\pi^{22}$ to $\cfg^n$, due to $(\ast)$. Then the diagram yields
$\pi^{pq}_{ij}(\tilde\partial C^{pqr}_{ijk}) \subseteq \cfg^n$.

Thus, $\cfg^n$, together with the $C^{pqr}_{ijk}$ of $p+q+r=n+1$ and the
corresponding restrictions of the $\pi^{pq}_{ij}$ taken as characteristic maps,
define a cell complex $\cfg^{n+1}$. We have to show that
$\cfg^{n+1}$ has the property $(\ast)$. For that purpose, let $\cfg^{n+1}_0$
denote the topological
direct sum of $\cfg^n$ with all the cells $C^{pqr}_{ijk}$ of dimension
$p+q+r=n+1$ and let $f$ denote the union of the attaching maps of these cells.
By definition, $\cfg^{n+1}$ is the quotient of $\cfg^{n+1}_0$ by the
equivalence relation
$$
x_1\sim_f x_2 ~~~\Leftrightarrow~~~ x_1=x_2 ~\text{ or }~ x_1 = f(x_2)
~\text{~or~}~ f(x_1) = x_2 ~\text{~or~}~ f(x_1) = f(x_2)\,,
$$
whenever $f(x_1) $ or $f(x_2)$ are defined. Define a map $\psi :
\cfg^{n+1}_0 \to \cfg$ to be the identity on $\cfg^n$ and
$\pi^{pq}_{ij}$ on $C^{pqr}_{ijk}$. Since the domain is compact and the target
space is Hausdorff, $\psi$ is closed, and hence induces a homeomorphism
between its image and the quotient of $\cfg^{n+1}_0$ obtained by contraction of
preimages. The image of $\psi$ can be easily found to be the image of the
$(n+1)$-skeleton of $\fdom^2\times\fdom^2\times D(\fdom^2,\fdom^2)$ under
$\pi^{22}$. Thus, all we have
to check is that $x_1\sim_f x_2$ iff $\psi(x_1) = \psi(x_2)$.

First, assume that $x_1\sim_f x_2$. If $x_1=x_2$ then
trivially $\psi(x_1) = \psi(x_2)$. If $x_1 = f(x_2)$ then
$x_1\in\cfg^n$ and $x_2$ belongs to the boundary of one of the
$C^{pqr}_{ijk}$. Then $x_1 = \psi(x_1)$ and
$f(x_2) = \pi^{pq}_{ij}(x_2) = \psi(x_2)$, hence $\psi(x_1) =
\psi(x_2)$, too. A similar argument applies to the cases $f(x_1) =
x_2$ and $f(x_1) = f(x_2)$.

For the converse implication, we need the following lemma.

\ble\label{LTCW}

Let $x\in C^{pqr}_{ijk}$ and assume that there exists $x'\in
C^{p'q'r'}_{i'j'k'}$, where $p+q+r \geq p'+q'+r'$ and $x'\neq x$, such that
$\pi^{pq}_{ij} (x) = \pi^{p'q'}_{i'j'} (x')$. Then $x\in\partial
C^{pqr}_{ijk}$.

\ele

According to Lemma \rref{Lvp}, $\pi^{pq}_{ij} (x) =
\pi^{p'q'}_{i'j'} (x')$ implies that $x$ and $x'$ have the same
$\fdom$-parts, i.e., $x = (a,b,y)$ and $x' = (a,b,y')$, where
$y\in D(\fdom^p_i,\fdom^q_j)$ and $y'\in D(\fdom^{p'}_{i'},\fdom^{q'}_{j'})$.
Assume that $x$ is in the interior of $C^{pqr}_{ijk}$. Then $a$ and $b$ are in
the interiors of $\fdomp_i$ and $\fdomq_j$ which are therefore intersected by
$\fdom^{p'}_{i'}$ and $\fdom^{q'}_{j'}$, respectively. Hence, \eqref{GLTCW1} holds and we
have the commutative diagram \eqref{Gdgr}, where the left vertical arrow is
again cellular by assumption. Hence, the image $\tilde y$ of $y'$ under this
map belongs to an $s$-cell of $D(\fdom^p_i,\fdom^q_j)$, where $s\leq r'$.
Due to the diagram, $\pi^{pq}_{ij}(a,b,\tilde y) = \pi^{p'q'}_{i'j'}(a,b,y')$
and hence $\pi^{pq}_{ij}(a,b,\tilde y) = \pi^{pq}_{ij}(a,b,y)$. Since
$\pi^ {pq}_{ij}$ is injective on $\intfdomp_i \times \intfdomq_j \times
D(\fdom^p_i,\fdom^q_j)$, then $\tilde y = y$. We conclude that $y$ belongs to
an $s$-cell of $D(\fdom^p_i,\fdom^q_j)$. Since $x$ is in the interior of
$C^{pqr}_{ijk}$, $y$ is in the interior of the corresponding $r$-cell of
$D(\fdom^p_i,\fdom^q_j)$, hence $s \geq r$ and so $r'\geq r$. Then, under the assumption $p+q+r
\geq p' + q' + r'$, \eqref{GLTCW1} implies $p' = p$ and $q' = q$
and, consequently, $i' = i$, $j'= j$. It follows $y' = \tilde y$, hence $y' = y$,
hence $x=x'$, in contradiction to the assumption. This proves the lemma.
\medskip

We continue with the proof of Theorem \rref{TCW}. Assume that
$\psi(x_1) = \psi(x_2)$. If $x_1 = x_2$ then $x_1\sim_f x_2$.
Hence, let us assume $x_1\neq x_2$. Then $x_1$ and $x_2$ cannot
both belong to $\cfg^n$, because $\psi$ is the identity there. If
one of them, say $x_1$, belongs to $\cfg^n$ and the other one to
one of the $n+1$-cells, say $x_2\in C^{pqr}_{ijk}$, then $\psi(x_2)
= x_1$. Of course,
% then $\pi^{pq}_{ij}(x_2) = x_1$. Of course,
$x_1$ belongs to the interior of some cell $C^{p'q'r'}_{i'j'k'}$ with
$p' + q' + r' \leq n$ and we have $x_1\equiv\pi^{p'q'}_{i'j'}(x_1)$, as we
have identified the interior of $C^{p'q'r'}_{i'j'k'}$ through
$\pi^{p'q'}_{i'j'}$ with its image in $\cfg$. Then Lemma \rref{LTCW} yields
$x_2\in\partial C^{pqr}_{ijk}$. It follows $f(x_2) =
\pi^{pq}_{ij}(x_2) = \psi(x_2)$, hence $f(x_2) = x_1$, i.e.,
$x_1\sim_f x_2$. If $x_1$ and $x_2$ both belong to one of the
$n+1$-cells $C^{pqr}_{ijk}$, then Lemma \rref{LTCW}, applied to
both $x_1$ and $x_2$, implies that they belong to the respective
boundaries. Therefore, as above, $f(x_l) = \psi(x_l)$, $l=1,2$. It
follows $f(x_1) = f(x_2)$, hence $x_1\sim_f x_2$, too.
\medskip

Finally, we determine the boundary operator. Let
$\fdom^p_i\times\fdom^q_j\times K$ be given. The boundary operator on the
level of the cell complex
$\fdom^p_i\times\fdom^q_j\times D(\fdom^p_i,\fdom^q_j)$ is given by
 \beq
 \label{GdAAK2}
\tilde\partial(\fdom^p_i\times\fdom^q_j\times K)
 =
\partial\fdom^p_i \times \fdom^q_j\times K
 +
(-1)^p \fdom^p_i\times \partial\fdom^q_j\times K
\\
 +
(-1)^{p+q}\fdom^p_i\times\fdom^q_j\times \partial K
 \,,
 \eeq
where the boundaries on the rhs.\ are taken in $\fdom$ and
$D(\fdom^p_i,\fdom^q_j)$, respectively. In order to obtain $\partial
(\fdom^p_i\times\fdom^q_j\times K)$ from this formula we have to replace each
cell $C$ appearing on the rhs.\ by the cells of $\cfg$
which span $\pi^{pq}_{ij}(C)$. Since the last term already consists of cells
of $\cfg$, it remains unchanged. The first term is a sum over cells of the type
$\fdom^{p\minus 1}_{i'}\times \fdom^q_j\times K$, where $\fdom^{p\minus 1}_{i'}$ is
one of the boundary cells of $\fdom^p_i$ (equipped with the correct sign).
We have
$$
\pi^{pq}_{ij}\Big(\fdom^{p-1}_{i'}\times \fdom^q_j\times K\Big)
 =
\pi^{p\minus 1~q}_{i'j}\Big(\fdom^{p-1}_{i'}\times \fdom^q_j\times
\rho_{i'}(K)\Big)\,,
$$
where $\rho_{i'}$ stands for the factorization map
$D(\fdom^p_i,\fdom^q_j) \to D(\fdom^{p\minus 1}_{i'},\fdom^q_j)$.
Since the argument of $\pi^{p\minus 1~q}_{i'j}$ on the rhs.\
consists of cells of $\cfg$, it replaces $\fdom^{p-1}_{i'}\times
\fdom^q_j\times K$ in \eqref{GdAAK2}, where
$\rho_{i'}$ has to be replaced by the induced homomorphism
$\rho_{i'\,\ast}$. Treating the 2nd term in \eqref{GdAAK2} in a similar way,
we obtain the asserted formula.
 \qed
\\

Next, we construct cell decompositions of the double quotients
\eqref{Gdoquos} which meet the assumptions of Theorem \rref{TCW}.
We shall start with $\TUT$.

%%%%%%%%%%%%%%%%%%%%%%%%%%%%%%%%%%%%%%%%%%%%%%%%%%%%%%%%%%%%%%%%%%%%%
%%%%%%%%%%%%%%%%%%%%%%%%%%%%%%%%%%%%%%%%%%%%%%%%%%%%%%%%%%%%%%%%%%%%%

\section{The double quotient $\TUT$}
\label{SY}
\setcounter{equation}{0}

%%%%%%%%%%%%%%%%%%%%%%%%%%%%%%%%%%%%%%%%%%%%%%%%%%%%%%%%%%%%%%%%%%%%%
%%%%%%%%%%%%%%%%%%%%%%%%%%%%%%%%%%%%%%%%%%%%%%%%%%%%%%%%%%%%%%%%%%%%%

Perhaps the most obvious way to treat the double quotient $\TUT$ is to
view it as the quotient of the left $T$-action on the flag manifold
$\mr U(3)/T$ and to construct a cell decomposition of this quotient
from the Schubert cells of $\mr U(3)/T$ associated with Borel subgroups
of $\GL(3,\CC)$ that contain $T$. However, we will not follow this road. Instead
of working with Schubert cells, we will relate the double quotient $\TUT$ with
the bistochastic and unistochastic $(3\times 3)$-matrices and define the cells
directly by conditions on the entries of the matrices they contain. The
relation between the cells so constructed and the Schubert cells of $\mr U(3)/T$
will be clarified in Appendix \rref{S-Bruhat}.

We start with introducing some notation. In this section we use the shorthand
notation $Y := \TUT$. Let $Q\subseteq Y$
denote the subset of classes that have real representatives, i.e.,
which intersect $\mr O(3)$. Let $\mc B_3$ denote the set of
$(3\times 3)$-matrices with real nonnegative entries that add up
to $1$ in each row and each column. Such a matrix is called
bistochastic. $\mc B_3$ has the structure of a convex polytope. It
is known as the Birkhoff polytope of rank $3$. The corners of this
polytope are given by the permutation matrices. There exists a
natural map $\psi : \mr U(3)\to \mc B_3$, given by
 \beq\label{G-defpsi}
\psi(a)_{ij} := |a_{ij}|^2\,.
 \eeq
A point in the image of $\psi$ is called a unistochastic matrix. A
point in the image of the restriction of $\psi$ to $\mr O(3)$ is
called an orthostochastic matrix. The subsets of unistochastic and
orthostochastic matrices are denoted by $\mc U_3$ and $\mc O_3$,
respectively. Bistochastic and unistochastic matrices have several
applications in mathematics, computer science and physics, see the
introduction of \cite{Birkhoff} for a brief overview. It is known
that $\mc U_3$ is a closed star-shaped $4$-dimensional subset of
$\mc B_3$ and that $\mc O_3$ is its boundary \cite[Thm.\
3]{Birkhoff}. Hence, topologically, $\mc U_3$ is a $4$-disk and
$\mc O_3$ is a $3$-sphere. To make this information available for
the study of $Y$, we observe that the map $\psi$ descends to a
continuous and surjective map (same notation)
$$
\psi : Y\to\mc U_3\,.
$$
This map will now be analyzed. For a $(3\times 3)$-matrix $a$, let
$\ol a$ denote the complex conjugate matrix, i.e., $(\ol a)_{ij} =
\ol{a_{ij}}$. Since $\ol T = T$, complex conjugation induces a
well-defined map $Y\to Y$ which will be denoted $y\mapsto \ol y$,
too. This map is a homeomorphism.

\ble\label{L-TUT-1}

Let $y\in Y$. Then $\ol y = y$ if and only if $y\in Q$.

\ele

{\it Proof.}~ We have to show that $\ol y = y$ implies $y\in Q$. Let $a\in\mr
U(3)$ be a representative of $y$. By assumption, there exist $b_1,b_2\in T$ such
that $\ol a = b_1 a b_2$. For each $b_i$ there exists $c_i\in T$ such that
$c_i^2 = b_i$. Then $\ol{c_1} \ol a \ol{c_2} = c_1 a c_2$, i.e., $c_1 a c_2 \in
\mr O(3)$ and hence $y\in Q$.
 \qed

\ble\label{L-TUT-2}

Let $y_1,y_2\in Y$. If $\psi(y_1) = \psi(y_2)$ then $y_2 = y_1$ or
$y_2 = \ol{y_1}$.
 \qed

\ele

{\it Proof.}~ Let $y\in Y$. Obviously, $\psi(\ol y) = \psi(y)$.
Hence, we have to show that $y$ and $\ol y$ are the only elements
of $Y$ that are mapped to $\psi(y)$ under $\psi$. 
The question to what extent a unitary matrix of rank $3$ is
determined by the moduli of its entries was discussed in
\cite{Jarlskog} in connection with $CP$-violation in the standard
electroweak model. The proof uses the unitarity triangles introduced there. 
Let $a$ be a representative of $y$. Up to the action of $T\times T$ we may
assume that the entries of the first row and the first column of $a$ are real
and nonnegative. Define complex numbers
$$
u_i = \ol{a_{1i}}a_{2i}
 \,,~~~~~~
u^i = \ol{a_{i1}}a_{i2} 
 \,,~~~~~~
v_i = \ol{a_{1i}}a_{3i}
 \,,~~~~~~
v^i = \ol{a_{i1}}a_{i3} 
 \,,~~~~~~
i,j=1,2,3\,.
$$
Unitarity implies $\sum_i u_i = \sum_i u^i = \sum_i v_i = \sum_i v^i = 0$.
Hence, the triples $(u_1,u_2,u_3)$, $(u^1,u^2,u^3)$, $(v_1,v_2,v_3)$,
$(v^1,v^2,v^3)$ form triangles in the complex
plane, with one side on the real axis. These triangles are called unitarity
triangles. They are possibly degenerated to a line. The following data of these
triangles are determined by $y$ and the choice of $a$: the side on the real
axis, the length of the other two sides, the order of sides. The crucial
observation is that these data fix the triangles up to complex conjugation,
i.e., up to the transformations
$(u_1,u_2,u_3)\mapsto(\ol{u_1},\ol{u_2},\ol{u_3})$ etc.~. Now, taking the
complex conjugate of $(u_1,u_2,u_3)$ requires taking the complex conjugate of 
the entries $a_{22}$ and $a_{23}$, hence implies taking the complex conjugate of
$(u^1,u^2,u^3)$. Iterating this argument we find that one has to take the 
complex conjugate of all the triangles, and hence of all the entries of $a$,
simultaneously. This proves the lemma.
 \qed

\ble\label{L-TUT-3}

The map $\psi$ is open.

\ele

{\it Proof.}~ Since $\psi$ maps from a Hausdorff space to a
compact space, it is closed. Since it is also surjective, it maps
open subsets that are saturated, i.e., consist of full pre-images,
to open subsets. Hence, it suffices to show that the saturation of
an open subset $M$ is open. By Lemma \rref{L-TUT-2}, the
saturation is given by $M\cup \ol M$. Since the map $y\mapsto \ol
y$ is a homeomorphism, $M\cup \ol M$ is open.
 \qed

\bpr\label{P-TUT-cov}

By restriction, $\psi$ induces a $2$-fold covering $Y\setminus Q
\to \mc U_3\setminus\mc O_3$ and a homeomorphism $Q\to\mc O_3$.

\epr

\bre

This $2$-fold covering carries in fact the structure of a locally
trivial principal fibre bundle with structure group $\ZZ_2$,
acting by conjugation.

\ere

{\it Proof.}~ By construction, the maps are well-defined and
surjective. According to Lemmas \rref{L-TUT-1} and \rref{L-TUT-2},
$\psi$ is injective on $Q$. By Lemma \rref{L-TUT-3}, it is then a
homeomorphism. To check that the restriction of $\psi$ to
$Y\setminus Q$ yields a $2$-fold covering, let $y\in Y\setminus Q$
and denote $u = \psi(y)$. By Lemma \rref{L-TUT-1}, $y\neq \ol y$.
Since $Y$ is Hausdorff, there exist disjoint open neighbourhoods
$V_1$ of $y$ and $V_2$ of $\ol y$. Let $V := V_1\cap \ol V_2$ and
$U := \psi(V)$. By construction, $V$ and $\ol V$ are disjoint open
neighbourhoods of $y$ and $\ol y$, respectively. By Lemma
\rref{L-TUT-2}, $\psi^{-1}(U) = V\cup \ol V$. Lemma \rref{L-TUT-3}
implies that $U$ is an open neighbourhood of $u$ and that, by
restriction, $\psi$ induces homeomorphisms $V$ and $\ol V$ onto
$U$.
 \qed
\\

Since $\mc U_3$ is a $4$-disk with boundary $\mc O_3$, $Y\setminus
Q$ consists of $2$ connected components, each of which is a copy of
the open $4$-disk. Denote these connected components by $Y_\pm$.
Since $Q$ is closed, the $Y_\pm$ are open in $Y$. According to
Proposition \rref{P-TUT-cov}, by restriction, $\psi$ induces a
homeomorphism of $Y_\pm\cup Q$ onto $\mc U_4$. In particular,
$Y_\pm$ is dense in $Y_\pm\cup Q$. As $Y_\pm\cup Q$ has complement
$Y_\mp$ in $Y$, it is closed. Hence, the closure $\ol{Y_\pm}$ of
$Y_\pm$ in $Y$ is given by
$$
\ol{Y_\pm} = Y_\pm \cup Q.
$$
Thus, $\ol{Y_\pm}$ is a $4$-disk whose boundary is given by $Q$.
The main conclusion we draw from this is that any cell
decomposition of $Q$ combines with the two $4$-cells $\ol{Y_\pm}$
to a cell decomposition of $Y$. We denote $K^4_\pm := \ol{Y_\pm}$.

\bre

In addition, it follows that $\TUT$ is homeomorphic to a
$4$-sphere. This information does not help however in the
construction of a cell decomposition that meets the requirement
that the factorization maps are cellular.

\ere

We will now construct a cell decomposition of $Q\cong \mc O_3$.
The construction is based on the observation that under the
factorization maps classes of permutation matrices in $\TUT$ get
identified with one another in a variety of patterns. Therefore,
we take the permutation matrices as the $0$-cells. Denote them by
$K^0_\tau := \{\tau\}$, $\tau\in\mc S_3$.
 \bigskip

{\it $1$-skeleton:}~ Define
$$
K^1_{ij} := \{ b\in\mc B_3 : b_{ij} = 1\}
 \,,~~~~~~
i,j=1,2,3\,.
$$
Parameterisations of these subsets are given by
$$
b_{\ip\,\jp} = b_{\im\,\jm} = t
 \,,~~~~~~
b_{\ip\,\jm} = b_{\im\,\jp} = 1-t
 \,,~~~~~~
t\in[0,1]\,.
$$
Explicitly, for $i=j=1$,
$$
b =
 \left[
 \begin{array}{ccc}
1 & 0 & 0
\\
0 & t & 1-t
\\
0 & 1-t & t
 \end{array}
 \right]
 \,,~~~~~~
t\in[0,1]\,.
$$
We read off:

1.~ $K^1_{ij}\subseteq\mc O_3$:~ A representing orthogonal matrix
$a$ for $b$ is given by $a_{ij} = \pm \sqrt{b_{ij}}$ with
appropriately chosen signs. E.g., for the case $i=j=1$,
$$
a =
 \left[
 \begin{array}{ccc}
1 & 0 & 0
\\
0 & \sqrt t & - \sqrt{1-t}
\\
0 & \sqrt{1-t} & \sqrt t
 \end{array}
 \right]
 \,,~~~~~~
t\in[0,1]\,.
$$

2.~ For given $i$, $j$, there are two permutations mapping $j$ to
$i$. An explicit calculation yields that these permutations are
given by $\tau_i\tau_j\tau_i$ and $\tau_j\tau_i$. The first one is
odd, the second one is even. $K^1_{ij}$ is the line in the vector
space of real $(3\times 3)$-matrices connecting the permutation
matrices that correspond to these two permutations. Hence, the
$K^1_{ij}$ are $1$-disks and
 \beq\label{G-dK1-set}
\partial K^1_{ij} = K^0_{\tau_i\tau_j\tau_i} \cup
K^0_{\tau_j\tau_i}\,.
 \eeq

3.~ $K^1_{ij}\cap K^1_{\ipm\jpm} = K^0_{\tau_+}$ where $\tau_+$ is
the even permutation mapping $j$ to $i$ and $K^1_{ij}\cap
K^1_{\ipm\jmp} = K^0_{\tau_-}$ where $\tau_-$ is the odd
permutation mapping $j$ to $i$. All the other intersections are
trivial.

Thus, the cells $K^0_\tau$ and $K^1_{ij}$ so constructed, together
with the obvious attaching maps, form a cell complex of dimension
$1$.
 \bigskip

{\it $2$-skeleton:}~ For $i,j=1,2,3$, define
$$
K^2_{ij} := \{b\in\mc U_3 : b_{ij}=0\}\,.
$$
To be definite, the following argument is given for $K^2_{11}$. It
easily carries over to the other $K^2_{ij}$.

\ble\label{L-TUT-2cell}

The map
 \beq\label{G-par-2cell}
[0,1]^2 \to \mc B_3
 \,,~~~
(s,t) \mapsto
 \left[\begin{array}{ccc}
0 & t & 1-t
\\
s & (1-s)(1-t) & (1-s)t
\\
1-s & s(1-t) & st
 \end{array}\right]
 \eeq
induces a homeomorphism of $[0,1]^2$ onto $K^2_{11}$.

\ele

{\it Proof.}~ The map is injective and closed. Hence, it suffices
to check that its image coincides with $K^2_{11}$. Denote the
image by $\mc I$. According to \cite{Au-Yeung,Jarlskog}, a
bistochastic $(3\times 3)$-matrix is unistochastic iff for two
arbitrarily chosen rows or columns the 'chain-links' condition is
satisfied. For the 1st and 2nd column this condition reads
$$
 \left|
\sqrt{b_{21}} \sqrt{b_{22}} - \sqrt{b_{31}} \sqrt{b_{32}}
 \right|
 \leq
\sqrt{b_{11}} \sqrt{b_{12}}
 \leq
\sqrt{b_{21}} \sqrt{b_{22}} + \sqrt{b_{31}} \sqrt{b_{32}}\,.
$$
In case $b_{11}=0$ this yields
 \beq\label{G-clcond}
b_{21} b_{22} = b_{31} b_{32}\,.
 \eeq
Since this condition is satisfied for the elements of $\mc I$,
$\mc I\subseteq \mc U_3$ and hence $\mc I \subseteq K^2_{11}$.
Conversely, let $b\in K^2_{11}$. Then
$$
b =
 \left[\begin{array}{ccc}
0 & t & 1-t
\\
s & b_{22} & b_{23}
\\
1-s & b_{32} & b_{33}
 \end{array}\right]
$$
for some $s,t\in[0,1]$. Since $K^2_{11}\subseteq \mc U_3$,
\eqref{G-clcond} holds. Hence, $s b_{22} = (1-s) b_{32}$. Together
with $t + b_{22} + b_{32} = 1$, this yields $b_{22} = (1-s)(1-t)$
and $b_{32} = s(1-t)$. Then $b_{32} = (1-s)t$ and $b_{33} = st$.
Hence, $K^2_{11} \subseteq \mc I$.
 \qed
\\

We deduce:

1.~ $K^2_{ij} \subseteq\mc O_3$. A representing orthogonal matrix
$a$ is given by $a_{ij} = \pm\sqrt{b_{ij}}$ with appropriately
chosen signs. E.g., for $i,j=1$, in the parameterisation
\eqref{G-par-2cell},
$$
a =
 \left[
 \begin{array}{ccc}
0 & \sqrt t & \sqrt{1-t}
\\
\sqrt s & \sqrt{(1-s)(1-t)} & -\sqrt{(1-s)t}
\\
\sqrt{1-s} & -\sqrt{s(1-t)} & -\sqrt{st}
 \end{array}
 \right]\,.
$$

2.~ The $K^2_{ij}$ are $2$-disks. For the case of $K^2_{11}$, the
boundary is obtained by setting $s=0,1$ or $t=0,1$ in
\eqref{G-par-2cell}. For the general case this yields
 \beq\label{G-dK2-set}
\partial K^2_{ij}
 =
K^1_{i\,\jp} \cup K^1_{i\,\jm} \cup K^1_{\ip\,j} \cup K^1_{\im\,j}
 \,.
 \eeq

3.~ The intersection of $K^2_{11}$ with any other $K^2_{ij}$
consists of elements of $K^2_{11}$ with two zero entries. From the
parameterisation \eqref{G-par-2cell} we see that these elements
must have $s=0,1$ or $t=0,1$. Hence, the intersection is a union
of 1-cells. It is obvious that this holds for any intersection of
two distinct $K^2_{ij}$.

Thus, the 2-cells $K^2_{ij}$, together with the 1-skeleton and the
obvious attaching maps, yield a cell complex of dimension $2$.
 \bigskip

\begin{figure}

\begin{center}
\psfrag{k00}{ $K^0_{\tau_0}$}%
\psfrag{k01}{ $K^0_{\tau_1}$}%
\psfrag{k02}{ $K^0_{\tau_2}$}%
\psfrag{k03}{ $K^0_{\tau_3}$}%
\psfrag{k04}{ $K^0_{\tau_4}$}%
\psfrag{k05}{ $K^0_{\tau_5}$}%
\psfrag{k111}{ $K^1_{11}$}%
\psfrag{k112}{ $K^1_{12}$}%
\psfrag{k113}{ $K^1_{13}$}%
\psfrag{k121}{ $K^1_{21}$}%
\psfrag{k122}{ $K^1_{22}$}%
\psfrag{k123}{ $K^1_{23}$}%
\psfrag{k131}{ $K^1_{31}$}%
\psfrag{k132}{ $K^1_{32}$}%
\psfrag{k133}{ $K^1_{33}$}%
\includegraphics[width=12cm]{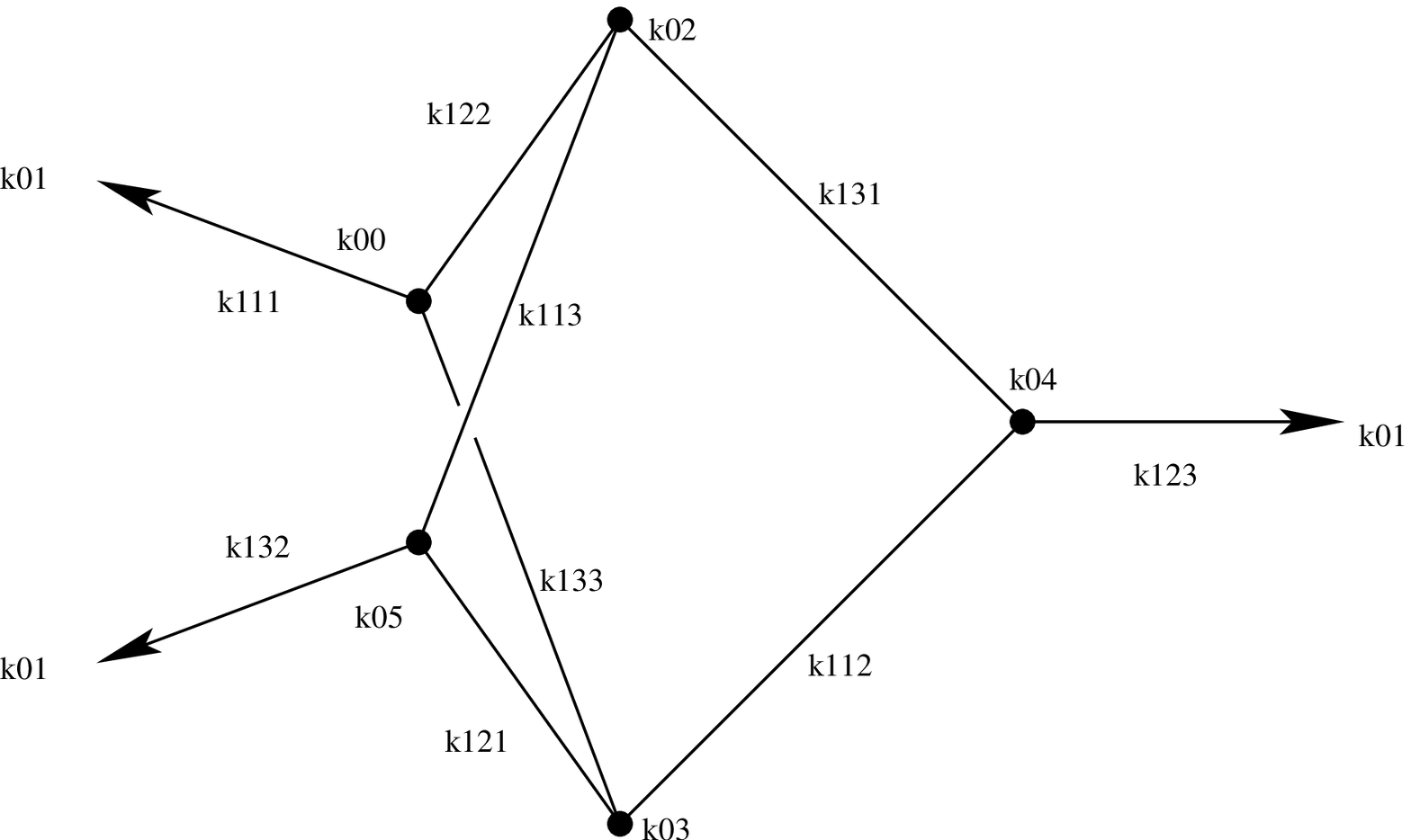}
\caption{\label{figureQ1} The subcomplex of the $2$-skeleton
consisting of cells that do not intersect $K^0_{\tau_1}$. The
$2$-cells form the faces of the triangular double pyramid.}

\end{center}

 \end{figure}

{\it $3$-skeleton:}~ Let $\tau\in\mc S_3$. Consider the subcomplex
of the $2$-skeleton consisting of cells that do not intersect
$K^0_\tau$. According to \eqref{G-dK1-set} and \eqref{G-dK2-set}
these cells are
 \beq\label{G-subcomplex}
K^0_\sigma\,,~~\sigma\neq\tau
 \,,~~~~~~
K^1_{ij}\,,~~ i\neq \tau(j)
 \,,~~~~~~
K^2_{ij}\,,~~ i = \tau(j)\,.
 \eeq
As $\mc O_3$ is a $3$-sphere, the subset $\mc O_3\setminus
K^0_\tau$ can be mapped homeomorphically onto $\RR^3$. This way,
the subcomplex \eqref{G-subcomplex} is embedded into $\RR^3$. A
simple inspection of the boundaries of the $1$ and $2$-cells then
shows that this subcomplex is homeomorphic to a $2$-sphere. As an
illustration, for the case of $\tau = \tau_1$ and for an
appropriately chosen homeomorphism $\mc O_3\setminus K^0_{\tau_1}
\cong \RR^3$, the $1$-skeleton of this subcomplex is shown in
Figure \rref{figureQ1}.

It follows that the subcomplex \eqref{G-subcomplex} cuts out a
subset of $\mc O_3$ homeomorphic to the $3$-disk. Denote this subset
by $K^3_\tau$. We take $K^3_\tau$, $\tau\in \mc S_3$, as the
$3$-cells. By construction,
 \beq\label{G-dK3-set}
\partial K^3_\tau = K^2_{\tau(1)\,1} \cup K^2_{\tau(2)\,2} \cup
K^2_{\tau(3)\,3}\,,
 \eeq
and the intersection of two distinct $3$-cells is a union of
$2$-cells. Thus, the $K^3_\tau$, together with the $2$-skeleton,
form a cell complex of dimension $3$. Since $\mc O_3$ coincides
with the union of the $3$-cells, we thus have constructed a cell
decomposition of $\mc O_3$. As stated above, together with
$K^4_\pm$, this yields a cell decomposition of $Y$.

\bre

The construction of cells is inspired by the polytope structure of
$\mc B_3$. The relation between the cells of $\mc O_3$ and the
faces of $\mc B_3$ is as follows. The $0$-cells of $\mc O_3$
coincide with the corners of $\mc B_3$. The $1$-cells of $\mc O_3$
are those edges of $\mc B_3$ that are contained in $\mc O_3$.
Since in $\mc B_3$ there is an edge between any two corners, there
are $6$ more edges in $\mc B_3$ that are not orthostochastic.
Since any $2$-face of $\mc B_3$ contains one of these
non-orthostochastic edges, the $2$-faces of $\mc B_3$ do not have
nontrivial intersection with $\mc O_3$. Instead, the $2$-cells of
$\mc O_3$ coincide with the intersections of the $3$-faces
('facets') of $\mc B_3$ with $\mc O_3$. The $3$-cells of $\mc O_3$
do not have an analogue in $\mc B_3$.

\ere

 \todo{

Remark on flag mf and Schubert-type cell decos

 }%

Next, we have to choose an orientation of the cells and to compute
the boundary operator.

\bpr\label{P-ori}

There exists an orientation of cells of $Y$ such that the boundary
operator is given by
 \begin{align*}
\partial K^1_{ij}
 & =
K^0_{\tau_i\tau_j\tau_i} - K^0_{\tau_j\tau_i}
 \,,~~~~~~
\partial K^2_{ij}
 =
K^1_{i\hjp} + K^1_{i\hjm} - K^1_{\hip j} - K^1_{\him j}
 =
\sum\nolimits_{l=1}^3 K^1_{il} - K^1_{lj}\,,
 \\
\partial K^3_\tau
 & =
\sign(\tau)~ \sum\nolimits_{i=1}^3 ~K^2_{\tau(i)\,i}
 \,,~~~~~~
\partial K^4_\pm
 =
\pm \sum\nolimits_{\tau\in\mc S_3} ~K^3_\tau\,.
 \end{align*}

\epr

{\it Proof.}~ We define the orientations inductively as follows.
For a cell in dimension $n+1$ we choose a certain boundary
$n$-cell and require the boundary orientation of this $n$-cell to
coincide with its genuine orientation chosen before.
 \bigskip

{\it $1$-cells:}~ Since according to \eqref{G-dK1-set}, each
$1$-cell connects $0$-cells of opposite sign, we can choose the
$0$-cells labelled by an odd permutation as starting points. This
yields the asserted formula for the $1$-cells.
 \bigskip

{\it $2$-cells:}~ Due to \eqref{G-dK2-set}, $K^2_{ij}$ is bounded
by four $1$-cells. Since neighbouring boundary $1$-cells intersect
in either their starting point or their end-point, they would
induce opposite orientations on $K^2_{ij}$. Hence, opposite
boundary $1$-cells induce the same orientation. Since there does
not exist a permutation mapping both $\jp$ and $\jm$ to $i$,
$K^1_{i\,\jp}$ and $K^1_{i\,\jm}$ are opposite. We choose the
orientation of $K^2_{ij}$ to be induced from these boundary
$1$-cells (i.e., those with coinciding first index). Since then
$K^1_{\ip\,j}$ and $K^1_{\im\,j}$ are opposite, too, the formula
for $\partial K^2_{ij}$ follows.
 \bigskip

{\it $3$-cells:}~ The boundary $2$-cells of $K^3_\tau$ are given
by \eqref{G-dK3-set}. Consider a $1$-cell $K^1_{kl}$ that belongs
to two of these boundary $2$-cells, say $K^2_{\tau(i)\,i}$ and
$K^2_{\tau(j)\,j}$, where $i\neq j$. Then $k=\tau(i)$ or $l=i$ and
$k=\tau(j)$ or $l=j$. Since $i\neq j$, then either $k=\tau(i)$ and
$l=j$ or $k=\tau(j)$ and $l=i$. In both cases, the boundary
orientations induced on $K^1_{kl}$ from $K^2_{\tau(i)\,i}$ and
$K^2_{\tau(j)\,j}$ are opposite. It follows that all three
boundary $2$-cells would induce the same orientation on
$K^3_\tau$. We could simply choose the orientations induced this
way for all the $3$-cells. However, these orientations would
obviously not combine to an orientation of the $3$-sphere $Q$,
because the orientations of intersecting $3$-cells would be
opposite. Instead, we choose the orientations as follows. Consider
a $2$-cell $K^2_{ij}$. By \eqref{G-dK3-set}, $K^2_{ij}$ belongs to
all $3$-cells labelled by a permutation that maps $j$ to $i$. As
noted above, these permutations are given by $\tau_j\tau_i$ and
$\tau_i\tau_j\tau_i$. Hence, each $2$-cell belongs to one $3$-cell
labelled by an even permutation and to one $3$-cell labelled by an
odd permutation. Thus, if we choose the orientation of $K^3_\tau$
to be induced from its boundary $2$-cells when $\tau$ is even and
to be opposite to that when $\tau$ is odd, these orientations
combine to an orientation of $Q$. The formula for the boundary
operator then follows.
 \bigskip

{\it $4$-cells:}~ By construction, all boundary $3$-cells of
$K^4_\pm$ would induce the same orientation on $K^4_\pm$. In order
that the orientations of $K^4_+$ and $K^4_-$ combine to an
orientation of $Y$, we choose that of $K^4_+$ to be induced from
its boundary $3$-cells and that of $K^4_-$ to be opposite. This
yields the asserted formula.
 \qed

 \bre

The $2$-skeleton can be conveniently visualized in a diagram, see
Figure \rref{hexagon}. In this diagram, six of the nine $2$-cells
have the shape of a trapezium and three of them form a rectangle
with one pair of opposite edges being twisted once. They are
labelled by symbols which mimic their shape.

\begin{figure}[h]

\begin{center}

\footnotesize

\unitlength3cm

\begin{picture}(0,3)

\put(0,0.3){
 \linie{0,0}{5,3}{1}
 \linie{0,0}{-5,3}{1}
 \linie{0,0}{5,3}{1}
 \linie{-1,0.6}{0,1}{1.2}
 \linie{1,0.6}{0,1}{1.2}
 \linie{-1,1.8}{5,3}{1}
 \linie{1,1.8}{-5,3}{1}
 \linie{0,0}{0,1}{2.4}
 \linie{-1,0.6}{5,3}{2}
 \linie{1,0.6}{-5,3}{2}
 \punkt{0,0}
 \punkt{-1,0.6}
 \punkt{1,0.6}
 \punkt{-1,1.8}
 \punkt{1,1.8}
 \punkt{0,2.4}
 \put(0,0){\circle{0.12}}
 \put(-1,1.8){\circle{0.12}}
 \put(1,1.8){\circle{0.12}}
 \marke{0.45,0.35}{tl}{K^1_{21}}
 \marke{-0.45,0.35}{tr}{K^1_{32}}
 \marke{0.25,2.17}{bl}{K^1_{31}}
 \marke{-0.45,2.05}{br}{K^1_{22}}
 \marke{-1,0.9}{cr}{K^1_{11}}
 \marke{1,1.2}{cl}{K^1_{12}}
 \marke{0,0}{tc}{K^0_{\tau_5}}
 \marke{-1,0.6}{tr}{K^0_{\tau_1}}
 \marke{-1,1.8}{br}{K^0_{\tau_0}}
 \marke{1,0.6}{tl}{K^0_{\tau_3}}
 \marke{1,1.8}{bl}{K^0_{\tau_4}}
 \marke{0,2.4}{bc}{K^0_{\tau_2}}
{ \linethickness{0.01mm}
% \linie{0.05,2.15}{1,0}{1.5}
 \marke{0,2.15}{cl}{K^1_{13}}
% \linie{-0.75,1.6}{-1,0}{0.75}
 \marke{-0.89,1.6}{bl}{K^1_{33}}
% \linie{-0.75,0.8}{-1,0}{0.75}
 \marke{-0.89,0.8}{tl}{K^1_{23}}
}
 \linie{0.4,0.3}{5,3}{0.2}
 \linie{0.4,0.3}{-5,3}{0.2}
 \linie{0.2,0.42}{5,3}{0.4}
 \vektor{0.6,0.42}{0,1}{0.24}
 \marke{0.45,0.46}{cc}{K^2_{22}}
 \linie{-0.4,0.3}{-5,3}{0.2}
 \linie{-0.4,0.3}{5,3}{0.2}
 \linie{-0.2,0.42}{-5,3}{0.4}
 \vektor{-0.6,0.42}{0,1}{0.24}
 \marke{-0.45,0.46}{cc}{K^2_{31}}
 \linie{0.4,2.1}{5,-3}{0.2}
 \linie{0.4,2.1}{-5,-3}{0.2}
 \linie{0.2,1.98}{5,-3}{0.4}
 \vektor{0.6,1.74}{0,1}{0.24}
 \marke{0.45,1.96}{cc}{K^2_{32}}
 \linie{-0.4,2.1}{-5,-3}{0.2}
 \linie{-0.4,2.1}{5,-3}{0.2}
 \linie{-0.2,1.98}{-5,-3}{0.4}
 \vektor{-0.6,1.74}{0,1}{0.24}
 \marke{-0.45,1.96}{cc}{K^2_{21}}
 \linie{0.95,1.08}{0,1}{0.24}
 \linie{0.95,1.08}{-5,-3}{0.2}
 \vektor{0.75,0.96}{0,1}{0.48}
 \linie{0.95,1.32}{-5,3}{0.2}
 \marke{0.85,1.2}{cc}{\,K^2_{11}}
 \linie{-0.95,1.08}{0,1}{0.24}
 \linie{-0.95,1.08}{5,-3}{0.2}
 \vektor{-0.75,0.96}{0,1}{0.48}
 \linie{-0.95,1.32}{5,3}{0.2}
 \marke{-0.85,1.2}{cc}{\;\!K^2_{12}}
 \linie{0.1,1.2}{5,3}{0.4}
 \linie{0.1,1.2}{5,-3}{0.4}
 \vektor{0.5,1.44}{0,-1}{0.48}
 \marke{0.361436,1.2}{cc}{K^2_{13}}
 \linie{-0.05,1.11}{-5,-3}{0.4}
 \vektor{-0.05,1.11}{0,-1}{0.48}
 \linie{-0.05,0.63}{-5,3}{0.4}
 \marke{-0.18856,0.87}{cc}{K^2_{33}}
 \linie{0.05,1.11}{5,-3}{0.4}
 \vektor{0.05,1.11}{0,-1}{0.48}
 \linie{0.05,0.63}{5,3}{0.4}
 \marke{0.18856,0.87}{cc}{K^2_{23}}
 \linie{-0.1,1.2}{-5,3}{0.4}
 \linie{-0.1,1.2}{-5,-3}{0.4}
 \vektor{-0.5,1.44}{0,-1}{0.48}
 \marke{-0.361436,1.2}{cc}{K^2_{13}}
 \linie{-0.05,1.29}{-5,3}{0.4}
 \vektor{-0.05,1.77}{0,-1}{0.48}
 \linie{-0.05,1.77}{-5,-3}{0.4}
 \marke{-0.18856,1.53}{cc}{K^2_{23}}
 \linie{0.05,1.29}{5,3}{0.4}
 \vektor{0.05,1.77}{0,-1}{0.48}
 \linie{0.05,1.77}{5,-3}{0.4}
 \marke{0.18856,1.53}{cc}{K^2_{33}}
}

\end{picture}

\caption{\label{hexagon}The $2$-skeleton of $Y$ with orientation.
The starting points of the $1$-cells are encircled. The
orientations of the $2$-cells are indicated by an arrow along the
boundary of the corresponding symbols. Note that the crossing
point at the center is not a vertex.}

\end{center}

\end{figure}

 \ere

 \todo{

Relation with diagram in Harada?

 }

%%%%%%%%%%%%%%%%%%%%%%%%%%%%%%%%%%%%%%%%%%%%%%%%%%%%%%%%%%%%%%%%%%%%%%%%%%%%%%%%%%%%%%%
%%%%%%%%%%%%%%%%%%%%%%%%%%%%%%%%%%%%%%%%%%%%%%%%%%%%%%%%%%%%%%%%%%%%%%%%%%%%%%%%%%%%%%%

\section{The other double quotients}
\label{Sother}
\setcounter{equation}{0}

%%%%%%%%%%%%%%%%%%%%%%%%%%%%%%%%%%%%%%%%%%%%%%%%%%%%%%%%%%%%%%%%%%%%%%%%%%%%%%%%%%%%%%%
%%%%%%%%%%%%%%%%%%%%%%%%%%%%%%%%%%%%%%%%%%%%%%%%%%%%%%%%%%%%%%%%%%%%%%%%%%%%%%%%%%%%%%%

We now turn to the description of the other double quotients and
the associated factorization maps. We use the following notation.
The standard $2$-simplex is denoted by $\sigma^2$. If we label
its vertices by $1,2,3$, the edges are given, in the standard
notation, by $[12]$, $[23]$ and $[31]$ and the vertices are given
by $[1]$, $[2]$, $[3]$. The edge $[ij]$ is oriented from $[i]$ to
$[j]$. Similarly, the standard $1$-simplex is denoted by
$\sigma^1$ and its vertices by $[1]$ and $[2]$. $\sigma^1$ is
oriented from $[1]$ to $[2]$. There is an obvious ambiguity in
this notation. However, whether $[1]$ and $[2]$ denote vertices of
$\sigma^1$ or $\sigma^2$ will always be clear from the context.

Define maps $\psi_i : \mc U_3 \to \sigma^2$, $\psi^i : \mc U_3 \to \sigma^2$ and
$\psi_{ij}:\mc U_3\to\sigma^1$ by
$$
\psi_i(b) = (b_{i1},b_{i2},b_{i3})
 \,,~~~~~~
\psi^i(b) = (b_{1i},b_{2i},b_{3i})
 \,,~~~~~~
\psi_{ij}(b) = (b_{ij},1-b_{ij})\,.
$$

\ble\label{L-UUT}

The maps $\psi_i\circ\psi$, $\psi^i\circ\psi$ and
$\psi_{ij}\circ\psi$ descend to homeomorphisms $\UiUT\to\sigma^2$,
$\TUUi\to\sigma^2$ and $\UiUUj\to\sigma^1$, respectively. If we
use these homeomorphisms to identify the quotients with the
corresponding simplices, the factorization maps \eqref{Gfactmaps}
satisfy
 \beq\label{G-UUT}
 \lambda^{12}_i = \psi_i \circ \psi
 \,,~~~~~~
\lambda^{21}_i = \psi^i \circ \psi
 \,,~~~~~~
\mu^{12}_{ij}\circ\lambda^{12}_i
 =
\mu^{21}_{ij}\circ\lambda^{21}_j
 =
\psi_{ij}\circ\psi\,.
 \eeq

\ele

{\it Proof.}~ First we have to check that the maps descend to the
quotients. For $\psi_i\circ\psi$, this follows from the
observation that the action of $U_i$ and $T$ on $\mr U(3)$ by left
multiplication do not differ in their effect on the $i$-th row of
a matrix. Similarly, for $\psi^i\circ\psi$, the action of $U_i$
and $T$ on $\mr U(3)$ by right multiplication do not differ in
their effect on the $i$-th column. Combining these two
observations we obtain the assertion for $\psi_{ij}\circ\psi$.
 \comment{
that the actions of $U_i\times U_j$ and $T\times T$ by left and
right multiplication do not differ in their effect on the element
$a_{ij}$.
 }%

Next, we show that the descended maps are 1:1. As maps from a
Hausdorff space to a compact space they are homeomorphisms then.
Surjectivity is obvious. To check injectivity of the map
$\UiUT\to\sigma^2$ induced by $\psi_i\circ\psi$, let $a,b\in\mr
U(3)$ such that $\psi_i\circ\psi(TaT) = \psi_i\circ\psi(TbT)$.
Then $b_{ij} = \alpha_j a_{ij}$ for some $\alpha_j\in\mr U(1)$,
$j=1,2,3$. Hence, up to the action of $T$ by right multiplication,
we may assume that the $i$-th rows of $a$ and $b$ coincide.
Consider $c:=ba^{-1}$. We have $b=ca$ and $c_{ii} = 1$, because
this is the scalar product of the $i$-th row of $b$ with the
$i$-th row of $a$. Hence, $c\in U_i$ and $a$ and $b$ define the
same class in the quotient $\UiUT$. For the map $\TUUi\to\sigma^2$
induced by $\psi^i\circ\psi$, the argument is analogous. For the
map $\UiUUj\to\sigma^1$ induced by $\psi_{ij}\circ\psi$, let
$a,b\in\mr U(3)$ such that $\psi_{ij}\circ\psi(TaT) =
\psi_{ij}\circ\psi(TbT)$. Then $b_{ij} = \alpha a_{ij}$ for some
$\alpha\in\mr U(1)$. By the action of $U_j$ by right
multiplication, the entry $a_{ij}$ can be multiplied by an
arbitrary phase and the other two entries in the $i$-th row of $a$
can be arbitrarily adjusted. Hence, up to this action, we may
assume that the $i$-th rows of $a$ and $b$ coincide. Then the same
argument as for $\psi_i\circ\psi$ yields that $a$ and $b$ belong
to the same class in the quotient $\UiUUj$.

Finally, the equalities \eqref{G-UUT} hold by construction.
 \qed
\\

\bre

A slightly different interpretation of the quotients $\TUUi$ and $\UiUT$ is
obtained as follows.
Extraction of the $i$-th row defines a map from $\mr U(3)$
to the $5$-sphere $\mr S^5$ that translates the action of $T$ on $\mr U(3)$ by 
left multiplication into the natural action of $T \cong \mr U(1)\times\mr
U(1)\times\mr U(1)$ on $\mr S^5\subseteq \CC^3$. 
This map descends to a homeomorphisms of $\mr U(3)/U_i$ onto complex projective
space $\CC\mr P^2$ that translates the action of $T$ on $\mr U(3)/U_i$ into the
action of $T$ on $\CC\mr P^2$ inherited from the natural action of $T$ on
$\CC^3$. Thus, the quotient $\TUUi$ may be identified with the quotient of
$\CC\mr P^2$ w.r.t.\ this action. A similar result holds for the quotient
$\UiUT$.

\ere

Now we use \eqref{G-UUT} to compute how the factorization maps
$\lambda^{12}_i$, $\lambda^{21}_i$, $\mu^{12}_{ij}$ and
$\mu^{12}_{ij}$ map the cells of $\TUT$ and $\sigma^2$,
respectively.

\bpr\label{P-factmaps}

The factorization maps \eqref{Gfactmaps} map the cells of $\TUT$
and $\sigma^2$ as follows:
$$
 \renewcommand{\arraystretch}{1.5}
 \begin{array}{c|c|c|c|c}
 & K^0_\tau & K^1_{jk} & K^2_{jk} & K^3_\tau, K^4_\pm
 \\ \hline
\lUT
 &
[\tau^{\minus 1}(i)]
 & %\renewcommand{\arraystretch}{1}\textstyle
 \begin{array}{rcl}
[k] & | & j=i \\{} [\kp\,\km] & | & j\neq i
 \end{array}
 & %\renewcommand{\arraystretch}{1}
 \begin{array}{rcl}
[\kp\,\km] & | & j = i
 \\
\sigma^2 & | & j\neq i
 \end{array}
 &
\sigma^2
 \\ \hline
\lTU
 &
[\tau(i)]
 & %\renewcommand{\arraystretch}{1}
 \begin{array}{rcl}
[j] & | & k=i
 \\{}
[\jp\,\jm] & | & k\neq i
 \end{array}
 & %\renewcommand{\arraystretch}{1}
 \begin{array}{rcl}
[\jp\,\jm] & | & k=i
 \\
\sigma^2 & | & k\neq i
 \end{array}
 &
\sigma^2
 \end{array}
$$

$$
\renewcommand{\arraystretch}{1.5}
 \begin{array}{c|c|c}
 & [k] & [k\,\kp]
 \\ \hline
\mUT
 &
 \begin{array}{rcl}
[1] & | & k = j \\{} [2] & | & k \neq j
 \end{array}
 &
 \begin{array}{rcl}
[2] & | & k = \jp \\ \sigma^1 & | & k \neq \jp
 \end{array}
 \\ \hline
\mTU
 &
 \begin{array}{rcl}
[1] & | & k = i \\{} [2] & | & k \neq i
 \end{array}
 &
 \begin{array}{rcl}
[2] & | & k = \ip \\ \sigma^1 & | & k \neq \ip
 \end{array}
 \end{array}
$$
Here $i,j,k = 1,2,3$ and $\tau\in\mc S_3$. In particular, these
maps are cellular. Their induced homomorphisms are given by
 \begin{align*}
{\lUT}{}_\ast (K^0_\tau) & = [\tau^{\minus 1}(i)]
 \,,~~ &
{\lUT}{}_\ast (K^1_{jk})
 & =
(\delta_{\him,j} - \delta_{\hip,j}) [\kp\,\km]
 \,,~~ &
{\lUT}{}_\ast (K^2_{jk})
 & =
(\delta_{\him,j} - \delta_{\hip,j})\sigma^2
 \,,
\\
{\lTU}{}_\ast (K^0_\tau) & = [\tau(i)]
 \,, &
{\lTU}{}_\ast (K^1_{jk})
 & =
(\delta_{\him,k} - \delta_{\hip,k})[\jp\,\jm]
 \,, &
{\lTU}{}_\ast (K^2_{jk})
 & =
(\delta_{\hip,k} - \delta_{\him,k})\sigma^2
 \,,
 \end{align*}
 \begin{align*}
{\mUT}{}_\ast ([k])
 & =
\delta_{jk}[1] + (1-\delta_{jk})[2]
 \,,~~ &
{\mUT}{}_\ast ([k\,\kp])
 & =
(\delta_{k\,j} - \delta_{k\,\hjm})\sigma^1
 \,,~~ &
\\
{\mTU}{}_\ast ([k])
 & =
\delta_{ik}[1] + (1-\delta_{ik})[2]
 \,,~~ &
{\mTU}{}_\ast ([k\,\kp])
 & =
(\delta_{k\,i} - \delta_{k\,\him})\sigma^1
 \,.
 \end{align*}

Cells not listed here are annihilated.

\epr

{\it Proof.}~ First, consider $\lambda^{12}_i$. For the $0$ and
$1$-cells, the formulae can be read off from the definition of the
cells and the definition of $\psi_i$. Due to $\psi(K^4_\pm) = \mc
U_3$, we have $\lambda^{12}_i(K^4_\pm) = \sigma^2$. To check the
formula for the $2$-cells, consider $\lambda^{12}_i(K^2_{jk}) =
\psi_i\circ\psi(K^2_{jk})$. From the parameterisation
\eqref{G-par-2cell} of $K^2_{11}$, carried over to $K^2_{jk}$, we
read off: if $i=j$ then the image coincides with the edge of
$\sigma^2$ where the $k$-th entry is zero, i.e., with
$[\kp\,\km]$. If $i\neq j$, the image is the whole of $\sigma^2$.
This yields the formula for the $2$-cells. The formula for the
$3$-cells then follows by observing that each of them contains a
$2$-cell with first index being different from $i$. For
$\lambda^{12}_i$, the argument is similar.

To find $\mu^{12}_{ij}$, for each cell of $\sigma^2$ we choose a
preimage under $\lambda^{12}_i$ and determine its image under
$\mu^{12}_{ij}\circ\lambda^{12}_{ij} = \psi_{ij}\circ\psi$. For
the vertex $[k]$, a preimage under $\lambda^{12}_i$ is given by
$K^0_\tau$ where $\tau$ obeys $\tau(k) = i$. The entries of the
corresponding permutation matrix are $\tau_{ij} = 1$ if $i =
\tau(j)$, hence if $j=k$, and $0$ otherwise. Hence,
$\psi_{ij}\circ\psi(K^0_\tau) = [1]$ if $j=k$ and $[2]$ otherwise.
For the edge $[k\,\kp]$, a preimage under $\lambda^{12}_i$ is
given by $K^2_{i\,\km}$. By $\psi_{ij}\circ\psi$, this subset is
mapped to $[2]$ if $j=\km$ and to $\sigma^1$ otherwise. Again, for
$\mu^{21}_{ij}$, the argument is similar.

Finally, consider the induced homomorphisms. They annihilate cells
that are mapped to lower-dimensional cells by the original
factorization maps. To determine the signs, we have to compare the
orientations of the cells and their images. For the $1$-cells this
can be done by finding out whether the starting point of the cell
is mapped to the starting point of the image or not. $2$-cells are only
relevant for $\TUT$. For a given $2$-cell we choose a boundary
$1$-cell from which the orientation is induced. Since the
$1$-cells of $\sigma^2$ carry the induced boundary orientation,
the $2$-cell acquires the same sign as this boundary $1$-cell.
 \qed

%%%%%%%%%%%%%%%%%%%%%%%%%%%%%%%%%%%%%%%%%%%%%%%%%%%%%%%%%%%%%%%%%%%%%
%%%%%%%%%%%%%%%%%%%%%%%%%%%%%%%%%%%%%%%%%%%%%%%%%%%%%%%%%%%%%%%%%%%%%

\section{Cell complex structure}
\label{SCW}

%%%%%%%%%%%%%%%%%%%%%%%%%%%%%%%%%%%%%%%%%%%%%%%%%%%%%%%%%%%%%%%%%%%%%
%%%%%%%%%%%%%%%%%%%%%%%%%%%%%%%%%%%%%%%%%%%%%%%%%%%%%%%%%%%%%%%%%%%%%

We now combine the pairs of cells of $\fdom$ with the cell
decompositions of the corresponding double quotients as prescribed
by Theorem \rref{TCW}. The list of cells, together with their
shorthand notation $C^{pqr}_{ijk}$ introduced in the proof of this
theorem, is given in the following table. The number of cells in
dimension $0,\dots,8$ is $9$, $18$, $24$, $27$, $24$, $15$, $9$,
$6$, $2$, respectively, their total number is $134$.
$$
 \renewcommand{\arraystretch}{1.4}\arraycolsep2pt
 \begin{array}{c|rclrclrcl}
\mbox{dim} & \multicolumn{9}{c}{\mbox{cells}~ (\tau\in\mc S_3
 \,,~ i,j=1,2,3\,,~a=1,2)}
\\ \hline
8 &~ C^{224}_\pm & = & \fdomll\times\fdomll\times K^4_\pm
\\
7 & C^{223}_\tau & = & \fdomll\times\fdomll\times K^3_\tau
\\
6 & C^{222}_{ij} & = & \fdomll\times\fdomll\times K^2_{ij}
\\
5 & C^{221}_{ij} & = & \fdomll\times\fdomll\times K^1_{ij}
 \,,~~ &
C^{212}_i & = & \fdomll\times\fdoml_i\times\sigma^2
 \,,~~ &
C^{122}_i & = & \fdoml_i\times\fdomll\times\sigma^2
\\
4 & C^{220}_\tau & = & \fdomll\times\fdomll\times K^0_\tau
 \,,~~ &
C^{211}_{ij} & = & \fdomll\times\fdoml_i\times [\jp\,\jm]
 \,,~~ &
C^{121}_{ij} & = & \fdoml_i\times\fdomll\times [\jp\,\jm]
\\
3 & C^{210}_{ij} & = & \fdomll\times\fdoml_i\times [j]
 \,,~~ &
C^{120}_{ij} & = & \fdoml_i\times\fdomll\times [j]
 \,,~~ &
C^{111}_{ij} & = & \fdoml_i\times\fdoml_j\times \sigma^1
\\
2 & C^{200}_i & = & \fdomll\times\fdomo_i\times \ast
 \,,~~ &
C^{020}_i & = & \fdomo_i\times\fdomll\times \ast
 \,,~~ &
C^{110}_{i\,j\,a} & = & \fdoml_i\times\fdoml_j\times [a]
\\
1 & C^{100}_{ij} & = & \fdoml_i\times\fdomo_j\times \ast
 \,,~~ &
C^{010}_{ij} & = & \fdomo_i\times\fdoml_j\times \ast
\\
0 & C^{000}_{ij} & = & \fdomo_i\times\fdomo_j\times \ast
\\
 \end{array}
$$

\btm\label{TCW2}

The cells $C^{pqr}_{ijk}$, together with the characteristic maps
induced by the projections $\pi^{pq}_{ij}$, see
\eqref{Gdefpipqij}, define a cell complex structure on $\cfg$. The
boundary operator is given by

\renewcommand{\baselinestretch}{1.6}\small\normalsize
$\partial C^{224}_\pm = \pm\sum_{m=0}^5 C^{223}_m$
 \,,~~~~~~
$\partial C^{223}_\tau = \sign(\tau)~\sum_{i=1}^3
C^{222}_{i\,\,\tau(i)}$\,,

$\partial C^{222}_{ij} = C^{221}_{i\jp} + C^{221}_{i\jm} -
C^{221}_{\ip\,j} - C^{221}_{\im\,j} + C^{122}_\hip -C^{122}_\him +
C^{212}_\hip - C^{212}_\him$\,,

$\partial C^{221}_{ij}
 =
C^{220}_{\tau_i\tau_j\tau_i} - C^{220}_{\tau_j\tau_i}
 +
C^{121}_{\ip\,j} - C^{121}_{\im\,j}
 +
C^{211}_{\jp\,\, i} - C^{211}_{\jm\,\, i}$\,,

$\partial C^{212}_i = - \sum_{j=1}^3 C^{211}_{ij}$
 \,,~~~~~~
$\partial C^{122}_i = - \sum_{j=1}^3 C^{121}_{ij}$
 \,,~~~~~~
$\partial C^{220}_\tau
 =
\sum_{i=1}^3 C^{210}_{i\,\,\tau(i)} +
C^{120}_{i\,\,\tau^{\minus1}(i)}$\,,

$\partial C^{211}_{ij} = C^{210}_{i\,\jp} - C^{210}_{i\,\jm} +
C^{111}_{\jp\, i} - C^{111}_{\jm\, i}$
  ~,~~~~
$\partial C^{121}_{ij} = C^{120}_{i\,\jp} - C^{120}_{i\,\jm} +
C^{111}_{i\,\jm} - C^{111}_{i\,\jp}$\,,

$\partial C^{210}_{ij} = C^{110}_{j\,i\,1} + C^{110}_{\jp\,\,i\,2}
+ C^{110}_{\jm\,\,i\,2} + C^{200}_\him - C^{200}_\hip$
 \,,~~~~~~
$\partial C^{120}_{ij} = - C^{110}_{i\,j\,1} -
C^{110}_{i\,\jp\,\,2} - C^{110}_{i\,\jm\,\,2} + C^{200}_\him -
C^{020}_\hip$\,,

$\partial C^{111}_{ij} = C^{110}_{i\,j\,2} - C^{110}_{i\,j\,1}$
 \,,~~~~~~
$\partial C^{110}_{i\,j\,a} = C^{100}_{i\,\jp} - C^{100}_{i\,\jm}
- C^{010}_{\ip\, j} + C^{010}_{\im\, j}$
 \,,~~~~~~
$\partial C^{200}_i = \sum_{j=1}^3 C^{100}_{ji}$\,,

$\partial C^{020}_i = \sum_{j=1}^3 C^{010}_{ij}$
 \,,~~~~~~
$\partial C^{100}_{ij} = C^{000}_{\im\, j} - C^{000}_{\ip\, j}$
 \,,~~~~~~
$\partial C^{010}_{ij} = C^{000}_{i\,\jm} - C^{000}_{i\,\jp}$\,.

\renewcommand{\baselinestretch}{1}\small\normalsize

\etm

{\it Proof.}~ The first statement is a reformulation of Theorem
\rref{TCW}. The formulae for the boundary map are a
straightforward consequence of the general formula given in the
theorem mentioned and the formulae for the factorization maps
given in Proposition \rref{P-factmaps}.
 \qed
 \\

%%%%%%%%%%%%%%%%%%%%%%%%%%%%%%%%%%%%%%%%%%%%%%%%%%%%%%%%%%%%%%%%%%%%%
%%%%%%%%%%%%%%%%%%%%%%%%%%%%%%%%%%%%%%%%%%%%%%%%%%%%%%%%%%%%%%%%%%%%%

\section{Stratification}
\label{Sstrat}
\setcounter{equation}{0}

%%%%%%%%%%%%%%%%%%%%%%%%%%%%%%%%%%%%%%%%%%%%%%%%%%%%%%%%%%%%%%%%%%%%%
%%%%%%%%%%%%%%%%%%%%%%%%%%%%%%%%%%%%%%%%%%%%%%%%%%%%%%%%%%%%%%%%%%%%%

In this section we show that the closures of the orbit type strata
are subcomplexes of $\cfg$ and determine the cells they consist
of. We start with introducing some notation and deriving some
formulae needed in the sequel.

Let $\tilde T := T\cap \SU(3)$ and $\tilde U_i := U_i\cap \SU(3)$.
We have $\tilde T \cong \mr U(1)\times\mr U(1)$ and $\tilde U_i
\cong \mr U(2)$. Let $T_i$, $i=1,2,3$ denote the subgroups
consisting of the matrices
$$
\diag(\ol\alpha^2,\alpha,\alpha)
 ~,~~~~
\diag(\alpha,\ol\alpha^2,\alpha)
 ~,~~~~
\diag(\alpha,\alpha,\ol\alpha^2)~,
$$
respectively, where $\alpha\in\mr U(1)$. The centralizers of the
subgroups $T_i$ and $\tilde U_i$ in $\SU(3)$ and $\U(3)$ are
$$
\mr C_{\U(3)}(T_i) = U_i
 ~,~~~~
\mr C_{\U(3)}(U_i) = T_i
 ~,~~~~
\mr C_{\SU(3)}(T_i) = \tilde U_i
 ~,~~~~
\mr C_{\SU(3)}(\tilde U_i) = T_i
$$
and their normalizers are
$$
\mr N_{\U(3)}(T_i) = \mr N_{\U(3)} (\tilde U_i) = U_i
 ~,~~~~
\mr N_{\SU(3)}(T_i) = \mr N_{\SU(3)} (\tilde U_i) = \tilde U_i\,.
$$
Furthermore, there holds
$$
\fdoml_i = \fdom\cap T_i\,.
$$
Since multiplication of an arbitrary matrix by a permutation
matrix from the left or the right results in the corresponding
permutation of rows or columns, respectively,
 \beq
 \label{GtU}
\tau U_i = U_{\tau(i)} \tau
 ~,~~~~
\tau\,\tilde U_i = \tilde U_{\tau(i)}\tau
 \,,~~~~~~
\tau T_i = T_{\tau(i)}\tau\,.
 \eeq
For $\tau\in\mc S_3$ and $i,j = 1,2,3$, define
$$
V^0_\tau := T\tau
 \,,~~~~~~
V^1_{ij} := \{a\in\mr U(3) : |a_{ij}| = 1\}
 \,,~~~~~~
V^2_{ij} := \{a\in\mr U(3) : a_{ij} = 0\}\,.
$$
The subsets $V^0_\tau$, $V^1_{ij}$ and $V^2_{ij}$ consist of the
representatives in $\mr U(3)$ of the elements of the cells
$K^0_\tau$, $K^1_{ij}$ and $K^2_{ij}$ of $\TUT$, respectively. We
have $V^1_{ii} = U_i$ and
 \beq\label{G-Vtau}
\tau V^r_{ij} = V^r_{\tau(i)\,j}
 \,,~~~~~~
V^r_{ij} \tau = V^r_{i\,\tau^{-1}(j)}
 \,,~~~~~~
r=1,2\,.
 \eeq

\ble\label{L-Vij}

~~~$ U_i\,\tau\,U_j =
 \begin{cases}
V^1_{ij} &|~ \tau(j) = i\,,
\\
V^2_{ij} &|~ \tau(j) \neq i\,.
 \end{cases}
 $

\ele

{\it Proof.}~ If $\tau(j)=i$, \eqref{GtU} and \eqref{G-Vtau} yield
$$
U_i\,\tau\,U_j = U_i \tau = V^1_{ii} \tau = V^1_{ij}\,.
$$
If $\tau(j)\neq i$, there exists a permutation $\sigma$ such that
$\sigma(1) = i$ and $\sigma(2) = \tau(j)$. Under the assumption
that there holds $U_1U_2 = V^2_{12}$, \eqref{GtU} and
\eqref{G-Vtau} imply
$$
U_i\,\tau\,U_j = \sigma\, U_1 U_2 \,\sigma^{-1}\tau = \sigma
\,V^2_{12} \,\sigma^{-1}\tau = V^2_{ij}\,.
$$
Hence, it suffices to show $U_1U_2 = V^2_{12}$. For any $a\in
U_1$, $b\in U_2$ we have $(ab)_{12} = \sum_{m=1}^3 a_{1m}b_{m2} =
0$, because $a_{1m}\neq 0$ only for $m=1$, whereas $b_{m2}\neq 0$
only for $m=2$. Hence, $U_1\,U_2\subseteq V^2_{12}$. To
prove the converse inclusion, let $a\in V^2_{12}$, i.e.,
$a_{12}=0$. Since the $2$nd column of $a$ is orthogonal to the
other two columns,
 \beq
 \label{GLVijk2}
\ol{a_{21}} a_{22} + \ol{a_{31}} a_{32} = 0
 ~,~~~~~~
\ol{a_{23}} a_{22} + \ol{a_{33}} a_{32} = 0~.
 \eeq
We view this as a system of linear equations in the variables
$a_{22}$ and $a_{32}$. Since $a$ does not have a zero row,
$a_{22}$ and $a_{32}$ cannot both vanish, so that this system has
a nontrivial solution. Hence, the determinant of the $(2\times
2)$-matrix
$$
\zzmatrix{a_{21}}{a_{23}}{a_{31}}{a_{33}}
$$
vanishes. Then this matrix is a tensor product, i.e., there exist
complex numbers $c_1,c_2,d_1,d_2$ such that $a_{21} = c_1d_1$,
$a_{23} = c_1d_2$, $a_{31} = c_2d_1$ and $a_{33} = c_2d_2$. We can
choose $c_1,c_2$ in such a way that $|c_1|^2+|c_2|^2 = 1$.
Consider the matrices
$$
a_1 = \ddmatrix{1}{0}{0}{0}{a_{22}}{c_1}{0}{a_{32}}{c_2}
 ,~~~~
a_2 = \ddmatrix{a_{11}}{0}{a_{13}}{0}{1}{0}{d_1}{0}{d_2}~.
$$
We have $a_1a_2 = a$. We check that $a_1$ is unitary: the columns
are obviously unit vectors. According to \eqref{GLVijk2},
$\ol{d_1}\left(\ol{c_1} a_{22} + \ol{c_2} a_{32}\right) = 0$ and
$\ol{d_2}\left(\ol{c_1} a_{22} + \ol{c_2} a_{32}\right) = 0$.
Since either $d_1\neq 0$ or $d_2\neq 0$, it follows that the 2nd
and 3rd row of $a_1$ are orthogonal, so that $a_1$ is unitary,
indeed. Then so is $a_2$. Thus, $a_1\in U_1$, $a_2\in U_2$, and
$a\in U_1U_2$, as asserted.
 \qed
\\

Now we turn to the discussion of the orbit type strata of $\cfg$.
The stabilizers of the action of $\SU(3)$ on $\SU(3)\times\SU(3)$
by diagonal conjugation are centralizers of pairs in $\SU(3)$. It
is well known that any subgroup of $\SU(3)$ which is a centralizer
is conjugate to one of the subgroups $\ZZ_3$, $T_1$, $\tilde T$,
$\tilde U_1$, $\SU(3)$. The corresponding orbit types will be
labelled by the numbers $1,\dots,5$ (in the respective order).
This numbering reflects the natural ordering of orbit types that
is inherited from the natural partial ordering of conjugacy
classes of subgroups of $\SU(3)$. I.e., type $n$ $\geq$ type $n'$
iff $n\geq n'$. The subset of $\cfg$ of orbits of type $n$ is
denoted by $\cfg_n$ and its closure in $\cfg$ by $\ol\cfg_n$. The
slice theorem for compact group actions \cite{Bredon:CTG} implies
that the orbit type subsets $\cfg_1,\dots,\cfg_5$ yield a disjoint
decomposition of $\cfg$ into manifolds satisfying the frontier
condition: if $\cfg_i\cap\ol{\cfg_j}$ then
$\cfg_i\subseteq\ol{\cfg_j}$. This orbit type decomposition is in
fact a stratification \cite{Pflaum:Strat}. Therefore we refer to
the subsets $\cfg_i$ as the orbit type strata of $\cfg$. Due to
the ordering of orbit types and the frontier condition,
$$
\ol\cfg_5\subseteq\cdots\subseteq\ol\cfg_1 = \cfg
 \,,~~~~~~
\ol\cfg_n = \bigcup\nolimits_{n'\geq n} \cfg_{n'}\,.
$$
In particular, $\cfg_1$ is the principal stratum.
\bigskip

\bre

Subgroups which can be written as a centralizer are 
called Howe subgroups and pairs of subgroups which centralize each
other are known as Howe dual pairs. Such pairs play a prominent
role in the representation theory of reductive Lie groups. For an
explicit listing of the Howe subgroups of the classical Lie groups
together with their partial ordering by inclusion modulo
conjugacy, see \cite{hdp}.

\ere

\ble\label{Lstrat}

For $(t,s,g)\in\fdom\times\fdom\times\mr U(3)$, the following
table lists the conditions on $g$ under which $\vp(t,s,g)$ belongs
to $\ol{\cfg_n}$, $n=2,\dots,5$.\rm
$$
\renewcommand{\arraystretch}{1.5}
\begin{array}{c||c|c|c|c}
\mbox{If } (t,s) \mbox{ is in the} &
\multicolumn{4}{c}{\vp(t,s,g)\mbox{ belongs to }}
 \\
\mbox{interior of}
 & \ol{\cfg_5} & \ol{\cfg_4} & \ol{\cfg_3} & \ol{\cfg_2}
 \\ \hline
\fdomll\times\fdomll
 & - & - &
g\in\bigcup_{\tau\in\mc S_3} V^0_\tau
 &
g\in\bigcup_{i,j=1}^3 V^1_{ij}
 \\ \hline
\fdoml_i\times\fdomll
 & - & - &
g\in\bigcup_{j=1}^3 V^1_{ij}
 &
g\in\bigcup_{j=1}^3 V^2_{ij}
 \\ \hline
\fdomll\times\fdoml_i
 & - & - &
g\in\bigcup_{j=1}^3 V^1_{ji}
 &
g\in\bigcup_{j=1}^3 V^2_{ji}
 \\ \hline
\fdoml_i\times\fdoml_j
 & - &
g\in V^1_{ij}
 &
g\in V^2_{ij}
 &
\mbox{all } g
 \\ \hline
\fdomo_i\times\fdomll\,,~\fdomll\times\fdomo_i
 & - & - &
\mbox{all } g
 & -
 \\ \hline
\fdomo_i\times\fdoml_j\,,~\fdoml_i\times\fdomo_j
 & - &
\mbox{all } g
 & - & -
 \\ \hline
\fdomo_i\times\fdomo_j
 &
\mbox{all } g
 & - & - & -
\end{array}
$$

\ele

{\it Proof.}~ The orbit type of $t,s,g$ is the conjugacy class of
the stabilizer of the pair $(t,gsg^{-1})$. The stabilizer is
$$
S(g) := \mr C_{\SU(3)}(t) \cap g C_{\SU(3)}(s) g^{-1}\,.
$$
$\fdomll\times\fdomll$:~ Let $(t,s)$ be in the interior of
$\fdomll\times\fdomll$. Then
$$
S(g) = \tilde T\cap g\tilde T g^{-1}\,.
$$
In particular, $S(g)$ is a Howe subgroup of $\SU(3)$ contained in
$\tilde T$. Hence, $S(g) = \tilde T$ (type $3$), $T_i$ (type $2$)
or $\ZZ_3$ (type $1$).

Case $S(g)  = \tilde T$:~ Here $g\in\mr N_{\U(3)}(\tilde T) = \mr
N_{\U(3)}(T) = \bigcup_{\tau\in\mc S_3} V^0_\tau$.

Case $S(g) = T_i$:~ Here $g^{-1}T_ig \subseteq \tilde T$, i.e.,
$g^{-1}T_ig$ is a Howe subgroup of $\SU(3)$ contained in $\tilde
T$. Hence, $g^{-1}T_ig = T_j$ for some $j=1,2,3$. Due to
\eqref{GtU}, $T_j = \tau^{\minus1}T_i\tau$, where $\tau\in\mc S_3$
is chosen so that $\tau(j) = i$. It follows $g\tau^{\minus1}\in\mr
N_{\U(3)}(T_i) = U_i$ and hence $g\in \bigcup_{\tau\in\mc S_3}
U_i\cdot\tau$. Conversely, if $g$ is of this form then $g^{-1} T_i
g = \tau^{\minus1}T_i\tau \subseteq\tau^{\minus1}\tilde T\tau =
\tilde T$, hence $T_i\subseteq S(g) $. Equality holds if $g$ is
not in $\bigcup_{\tau\in\mc S_3} V^0_\tau$. Due to \eqref{G-Vtau},
$\bigcup_{\tau\in\mc S_3} U_i\cdot\tau = \bigcup_{\tau\in\mc S_3}
V^1_{ii}\cdot\tau  = \bigcup_{j=1}^3 V^1_{ij}$. Since all $T_i$
belong to orbit type $2$, we have to take the union over
$i=1,2,3$, too.
\medskip

$\fdoml_i\times\fdomll$:~ Let $(t,s)$ be in the interior of
$\fdoml_i\times\fdomll$. Then
$$
S(g) = \tilde U_i\cap g\tilde T g^{-1}\,.
$$
This is a Howe subgroup of $\SU(3)$ contained in the maximal toral
subgroup $g\tilde T g^{-1}$. Hence, it can be $g\tilde T g^{-1}$
(type $3$), $gT_j g^{-1}$, $j=1,2,3$, (type $2$) or $\ZZ_3$ (type
$1$).

Case $S(g) = g\tilde T g^{-1}$:~ Under this assumption both
$g\tilde T g^{-1}$ and $\tilde T$ are maximal toral subgroups of
$\tilde U_i$, hence are conjugate in $\tilde U_i$. I.e., there
exists $h\in\tilde U_i\subseteq U_i$ such that $g \tilde T g^{-1}
= h\tilde T h^{-1}$. Then $h^{-1} g \in \mr N_{\U(3)}(\tilde T) =
N_{\U(3)}(T)$, hence $g\in U_i\cdot N_{\U(3)}(\tilde T) =
\bigcup_{\tau\in\mc S_3} U_i\cdot\tau = \bigcup_{j=1}^3 V^1_{ij}$.
The converse assertion is obvious.

Case $S(g) = gT_j g^{-1}$:~ Here $gT_j g^{-1}$ is contained in
some maximal toral subgroup of $\tilde U_i$. Hence, there exists
$h\in\tilde U_i\subseteq U_i$ such that $h^{-1} g T_j g^{-1} h
\subseteq \tilde T$. By the same argument as in the case of
$\fdom^2\times\fdom^2$ we conclude
that $g^{-1} h \in \bigcup_{\tau\in\mc S_3}U_j\cdot\tau$. Then
 \beq
 \label{GLstrat1}
g\in \bigcup\nolimits_{\tau\in\mc S_3} U_i\cdot \tau\cdot U_j\,.
 \eeq
Conversely, if $g$ is of this form, $g T_j g^{-1} \subseteq \tilde
U_i$, hence it is contained in $S(g)$, and is equal to $S(g)$ if
$g$ is not in $\bigcup_{k=1}^3 V^1_{ik}$. Since all values of $j$
belong to type $2$, in \eqref{GLstrat1} we have to take the union
over $j=1,2,3$. According to Lemma \rref{L-Vij}, this yields
$\bigcup_{j=1}^3 \bigcup_{\tau\in\mc S_3} U_i\cdot \tau\cdot U_j =
\bigcup_{j=1}^3 V^2_{ij}$.

For $(t,s)$ in the interior of $\fdomll\times\fdoml_i$, the proof
is analogous.
\medskip

$\fdoml_i\times\fdoml_j$:~ Let $(t,s)$ be in the interior of
$\fdoml_i\times\fdoml_j$. Then
$$
S(g) = \tilde U_i\cap g\tilde U_j g^{-1}\,,
$$
i.e., $S(g) $ is a Howe subgroup of $\SU(3)$ contained in $\tilde
U_i$. Hence, we can have $S(g) = \tilde U_i$, $S(g)
\cong\U(1)\times\U(1)$, $S(g) \cong\U(1)$ or $S(g) = \ZZ_3$.

Case $S(g) = \tilde U_i$:~ Let $\tau\in\mc S_3$ such that $\tau(j)
= i$. According to \eqref{GtU}, $g\tilde U_j g^{-1} = \tilde U_i =
\tau\tilde U_j\tau^{\minus 1}$. It follows $\tau^{\minus 1}
g\in\mr N_{\U(3)}(\tilde U_j) = U_j$ and thus $g\in\tau U_j$. The
converse implication is obvious. Due to \eqref{G-Vtau}, $\tau U_j
= \tau V_{jj} = V_{ij}$.

Case $S(g) \cong\U(1)\times\U(1)$:~ Here, $S(g) $ is a maximal
toral subgroup in $\SU(2)$. There exists $h\in\tilde U_i\subseteq
U_i$ such that $S(g) = h\tilde T h^{-1}$. It follows $\tilde
T\subseteq h^{-1}g\tilde U_j g^{-1} h$. By taking the centralizer
in $\SU(3)$ we obtain $h^{-1} g T_j g^{-1} h\subseteq\tilde T$. As
explained for the case of $\fdoml_i\times\fdomll$ above, then
$g\in\bigcup_{\tau\in\mc S_3} U_i\cdot\tau\cdot U_j$. Conversely,
if $g$ is of this form then $g\tilde U_j g^{-1} = h\tilde U_k
h^{-1}$ for some $k=1,2,3$ and $h\in U_i$. Hence,
$$
S(g) = \tilde U_i\cap h\tilde U_k h^{-1} = h(\tilde U_i\cap\tilde
U_k)h^{-1}
 \supseteq
h\tilde T h^{-1}\,,
$$
i.e., $S(g) $ contains a subgroup that is isomorphic to
$\U(1)\times\U(1)$. Equality holds if $g$ is not in $V^1_{ij}$.
Finally, Lemma \rref{L-Vij} yields $\bigcup_{\tau\in\mc S_3}
U_i\tau U_j = V^2_{ij}$.

Case $S(g) \cong\U(1)$:~ For any $t\in\fdoml_i$, $s\in\fdoml_j$,
both $t$ and $gsg^{-1}$ have a degenerate eigenvalue. The
intersection of the corresponding eigenspaces contains a
nontrivial common eigenvector $u$. Then the pair $(t,gsg^{-1})$ is
invariant under the $\U(1)$-subgroup of $\SU(3)$ defined by
multiplying $u$ by $\alpha^2$ and vectors orthogonal to $u$ by
$\ol\alpha$. It follows that the case $S(g) = \ZZ_3$ does not
occur, so that for all remaining $g$ the type of $\vp(t,s,g)$ is
$2$.
\medskip

Remaining cases:~ If $t\in\ZZ_3$, the orbit type of $\vp(t,s,g)$
is given by the centralizer of $s$, and vice versa.
 \qed

\btm\label{Tstrat}

The closures of the nonprincipal strata are subcomplexes of
$\cfg$. Their dimensions and the cells they consist of are listed
in the following table:
 \begin{center}
\rm
\renewcommand{\arraystretch}{1.3}
 \begin{tabular}{c|c|l}
 & $\dim$ & cells
\\ \hline
$\ol\cfg_5$ & $0$ & $C^{000}_{ij}$, $i,j=1,2,3$
\\
$\ol\cfg_4$ & $2$ & cells of $\ol\cfg_5$, ~$C^{100}_{ij}$,
$C^{010}_{ij}$, $C^{110}_{ij1}$, $i,j=1,2,3$
\\
$\ol\cfg_3$ & $4$ & cells of $\ol\cfg_4$, ~$C^{200}_i$,
$C^{020}_i$, $C^{110}_{ij2}$, $C^{210}_{ij}$, $C^{120}_{ij}$,
$i,j=1,2,3$, ~$C^{220}_\tau$, $\tau\in\mc S_3$
\\
$\ol\cfg_2$ & $5$ & cells of $\ol\cfg_3$, ~$C^{111}_{ij}$,
$C^{211}_{ij}$, $C^{121}_{ij}$, $C^{221}_{ij}$, $i,j=1,2,3$
 \end{tabular}
 \end{center}

\etm

A table with the numbers of cells of the closures of the strata in
each dimension is given in Appendix \rref{Snr}.
\\

{\it Proof.}~ We show that that the closures $\ol{\cfg_n}$ consist
of the asserted cells. That they are subcomplexes is then a
consequence of being closed and of being a union of cells.
 \comment{

wozu?

That the list of cells is complete, i.e., that with a cell it also
contains all its boundary cells, is checked for each $n=2,3,4,5$
using the formulae for the boundary operator given in Theorem
\rref{TCW2}.
 }%
Since $\ol\cfg_5\subseteq\cdots\subseteq\ol\cfg_2$, it is
convenient to go from $n=5$ downwards. Then for each $n$ we only
have to list cells that are not yet contained in $\ol\cfg_{n+1}$.
The subsets $\ol\cfg_n\subseteq\cfg$ are determined by means of
Lemma \rref{Lstrat} as a union over contributions of the kind
$\vp(\fdomp_i\times\fdomq_j\times V)$, where $V$ stands for
$V^0_\tau$, $V^1_{ij}$ or $V^2_{ij}$. Since on passing to $\TUT$,
the subsets $V^0_\tau$, $V^1_{ij}$ and $V^2_{ij}$ pass to the
cells $K^0_\tau$, $K^1_{ij}$ and $K^2_{ij}$, respectively, we have
$$
\vp(\fdomp_i\times\fdomq_j\times V) =
\pi^{pq}_{ij}(\fdomp_i\times\fdomq_j\times C)\,,
$$
where $C$ is a cell of the double quotient $D(\fdomp_i,\fdomq_j)$
which is obtained from the cell $K^0_\tau$, $K^1_{ij}$ or
$K^2_{ij}$ to which $V$ projects in $\TUT$ by application of the
appropriate factorization map \eqref{Gfactmaps}. We will explain
this for $\ol{\cfg_5}$ and $\ol{\cfg_4}$ and leave the rest to the
reader. For $\ol\cfg_5$, the lemma yields $\ol\cfg_5 =
\bigcup_{i,j=1}^3 \vp\big(\fdomo_i\times\fdomo_j\times \mr
U(3)\big)$. Here, the double quotient is trivial, hence this
coincides with
$$
\bigcup_{i,j=1}^3 \pi^{00}_{ij}\big(\fdomo_i\times\fdomo_j\times
\mr U(3)\big)
 =
\bigcup_{i,j=1}^3 \pi^{00}_{ij}\big(C^{000}_{ij}\big)\,.
$$
For $\ol\cfg_4$, from the lemma we read off
$$
\ol\cfg_4
 ~=~
 \bigcup\nolimits_{i,j=1}^3 ~ \Big(~
\vp\big(\fdoml_i\times\fdomo_j\times \mr U(3)\big)
 ~\cup~
\vp\big(\fdomo_i\times\fdoml_j\times \mr U(3)\big)
 ~\cup~
\vp\big(\fdoml_i\times\fdoml_j\times V^1_{ij}\big) ~\Big)\,.
$$
For the first two terms, the double quotient is again trivial,
hence these terms yield $\pi^{10}_{ij}(C^{100}_{ij})$ and
$\pi^{01}_{ij}(C^{010}_{ij})$. For the third term, the double
quotient is $\UiUUj$ and the cell to which $V^1_{ij}$ projects is
therefore
$$
\mUT\circ\lUT(K^1_{ij}) = [1]\,.
$$
Thus, the third term yields
$\pi^{11}_{ij}\left(C^{110}_{ij1}\right)$.
 \qed

\bre

\begin{figure}[hbt]
  \centering
  \epsfig{file=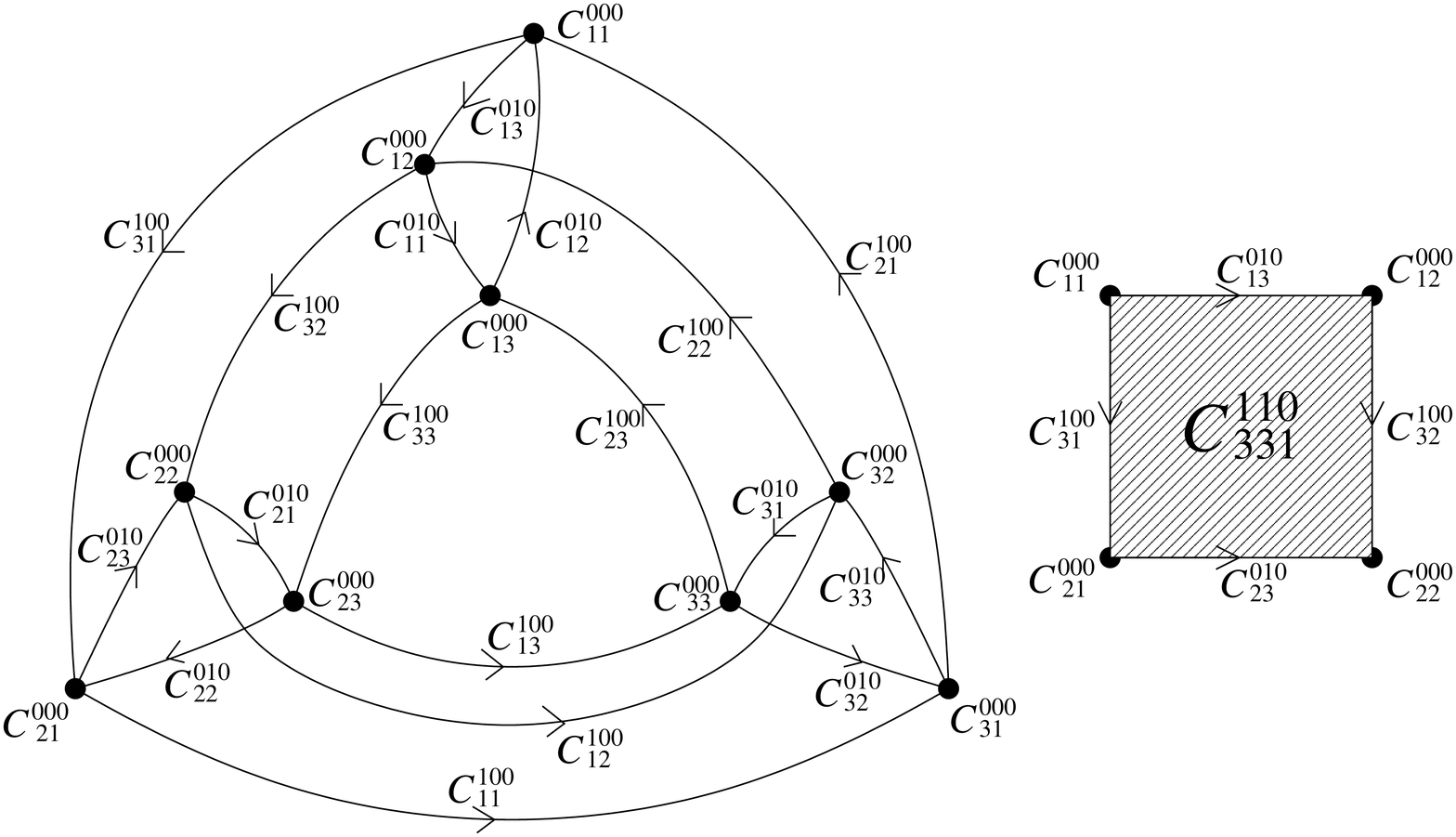,angle=0,width=0.8\textwidth}
  \caption{\label{2-stratum}The subcomplex $\ol\cfg_2$. The $2$-cells
  $C^{110}_{ij1}$ are not labelled. As an example, $C^{110}_{331}$ is pictured
  to the right.}

 \end{figure}

In a previous work \cite{confspace} it was shown that $\ol\cfg_4$
can be identified with $T_1\times T_1$, i.e., with a $2$-torus,
where the discrete subset $\ZZ_3\times\ZZ_3$ corresponds to
$\cfg_5$. Indeed, reading off from the boundary operator how the
cells of $\ol\cfg_2$ are attached to one another one can easily
see that they add up to a $2$-torus, see Figure \rref{2-stratum}.

\ere

%%%%%%%%%%%%%%%%%%%%%%%%%%%%%%%%%%%%%%%%%%%%%%%%%%%%%%%%%%%%%%%%%%%%%%%%%%%%%%%%%%%%%%%
%%%%%%%%%%%%%%%%%%%%%%%%%%%%%%%%%%%%%%%%%%%%%%%%%%%%%%%%%%%%%%%%%%%%%%%%%%%%%%%%%%%%%%%

\section{Homology and cohomology groups}
\label{Shml}
\setcounter{equation}{0}

%%%%%%%%%%%%%%%%%%%%%%%%%%%%%%%%%%%%%%%%%%%%%%%%%%%%%%%%%%%%%%%%%%%%%%%%%%%%%%%%%%%%%%%
%%%%%%%%%%%%%%%%%%%%%%%%%%%%%%%%%%%%%%%%%%%%%%%%%%%%%%%%%%%%%%%%%%%%%%%%%%%%%%%%%%%%%%%

%\subsection{Boundary homomorphisms and sub-complexes}

Let us start with recalling some basic facts. The homology groups
$H_k(K)$ of an $N$-dimensional cell complex $K$ are given by
$$
H_k(K):=\ker\partial_{k}/\image \partial_{k+1}\,,
$$
where $\partial_k$ is the boundary homomorphism in dimension $k$
of the chain complex
$$
0\longrightarrow
C_N(K)\mathop{\longrightarrow}^{\partial_N}C_{N-1}(K)
\mathop{\longrightarrow}^{\partial_{N-1}}\,\,\cdots\,\,
\mathop{\longrightarrow}^{\partial_1}C_0(K) \longrightarrow 0
$$
made up by the free abelian groups $C_k(K)$ based on the
$k$-dimensional cells of $K$. The homology groups $H_k(K,A)$ of
$K$ relative to the subcomplex $A$ are given by
$$
H_k(K,A):=\ker\tilde\partial_{k}/\image\tilde\partial_{k+1}\,,
$$
where $\tilde\partial_k$ is the boundary map in dimension $k$ of
the chain complex
 \beq
 \label{Gchain}
0\longrightarrow
C_N(K,A)\mathop{\longrightarrow}^{\tilde\partial_N}C_{N-1}(K,A)
\mathop{\longrightarrow}^{\tilde\partial_{N-1}}\,\,\cdots\,\,
\mathop{\longrightarrow}^{\tilde\partial_1}C_0(K,A) \longrightarrow 0\,.
 \eeq
Here $C_k(K,A) = C_k(K)/C_k(A)$ can be identified with the free
abelian group based on the $k$-cells of $K$ not in $A$ and
$\tilde\partial_k$ can be identified with $\partial_k$ composed
with projection to $C_{k-1}(K,A)$.
 \comment{Hence,
$C_k(K,A)\simeq\mathbb{Z}^{n_k}$, where $n_k$ is the number of
$k$-cells in $K$ which do not belong to $A$.
 }%
The boundary homomorphisms $\partial_k$ resp.\ $\tilde\partial_k$
can be represented by matrices $D_k$ by numbering the cells in
each dimension in an arbitrary way and defining the entry
$(D_k)_{ij}$ to be the coefficient with which the $j$-th
$(k-1)$-cell contributes to the image of the $i$-th $k$-cell under
$\partial_k$ resp.\ $\tilde\partial_k$ (the 'incidence number' of
these cells). By construction, the matrices $D_k$ have integer
entries and obey $D_kD_{k+1}=0$. The problem of computing the
homology groups is thus reformulated as the problem of computing $\ker
D_k/\image D_{k+1}$. To solve this, we will apply the following algorithm.
Recall that a finitely generated free
abelian group $G$ is isomorphic to $\Z^n$. Suppose that $H\subset
G$ is a subgroup. Then there exists a basis $e_1,\ldots,e_n$ of
$G$ and nonzero integers $q_1,\cdots,q_m$, $m\leqslant n$, such
that $q_i$ divides $q_{i+1}$ and $q_1e_1,\ldots,q_me_m$ is a basis
for $H$. In particular, $H$ is free abelian of rank $m$ and
 \beq
 \label{abeliangroup}
G/H\simeq\Z^{n-m}\oplus\Z_{q_1}\oplus\cdots\oplus\Z_{q_m}.
 \eeq
In our case, $G=\ker D_k$ and $H=\image D_{k+1}$. In order to find
the numbers $m$ and $q_i$ we can proceed as follows. There exist
unimodular matrices $U_k$ and $V_k$ such that $S_k := U_k D_k V_k$ is of the
so-called Smith normal form \cite{smith}. I.e.,

(i)~ $(S_k)_{ij} = 0$ for $i\neq j$ (we will sloppily say that $S_k$ is
diagonal, although it may not be square)

(ii)~ $(S_k)_{ii} > 0$ and $(S_k)_{ii}$ divides $S_{i+1~i+1}$ for
$i=1,\dots,r$,

(iii)~ $(S_k)_{ii} = 0$ for $i = r+1,\dots,n$.

We remark that $(S_k)$ is uniquely determined by $D_k$, whereas $U_k$
and $V_k$ are not. The columns of the matrix $V_k$ form a basis of the
domain $\mathbb{Z}^{n_k}$ of $D_k$ which has the property that the last
$n_k-r$ elements $v_{r+1},\ldots,v_{n_k}$ span $\ker D_k$. Therefore,
$\ker D_k\cong\mathbb{Z}^{n_k-r}$. Recall that the dimension $n_k-r$ can be
read off from $S_k$ as the number of zero columns. The submatrix
$$
P_k^0:=\left\{(V_k^{-1})^i_j\right\}^{i=r+1\ldots n_k}_{j=1\ldots n_k}
$$
of $V_k^{-1}$ yields the projection $\mathbb{Z}^{n_k}\to \ker
D_k$ which corresponds to the decomposition defined by the
basis $v_i$. We define $D_{k+1}^0:=P_k^0 D_{k+1} : \ZZ^{n_{k+1}}\to \ker D_k$.
As $\image D_{n_k+1}\subseteq\ker D_k$, we have $\image D_{n_k+1} = \image
D_{n_k+1}^0$ and hence $H_k=\ker D_k/\image D_{k+1}^0 \cong \ZZ^{n_k-r}/\image
S^0_{k+1}$, where $S_{k+1}^0$ denotes the Smith normal form of $D^0_{k+1}$.
Since $S^0$ is diagonal, we can analyze the factorization in each component
independently. Every row equal to zero in $S^0$ corresponds to a
generator in $\mathbb{Z}^{n_k-r}$ which does not appear in $\image
S^0$ and hence generates a factor $\Z$ in the quotient
$\mathbb{Z}^{n_k-r}/\image S^0$. Each nonzero diagonal entry $q$ contributes a
finite cyclic factor $\ZZ_q = \ZZ/ q \ZZ$ to the quotient,
where of course the factors with $q=1$ can be omitted. Summarizing,
$$
H_k = \mathbb{Z}^l\oplus_{q}\mathbb{Z}_q\,,
$$
where $l$ is the number of zero rows in $S^0$ and $q$ runs through the diagonal
entries distinct from $0$ and $1$.

Let us illustrate the procedure with the following example. Consider the group
$H_4(\ol\cfg_2)$. Due to Theorem \ref{Tstrat}, the cell complex
$\ol\cfg_2$ has 9, 24 and 27 cells in dimension 5, 4 and 3 respectively.
Hence, the corresponding part of the chain complex \eqref{Gchain} reads
$$
\mathbb{Z}^9\mathop{\longrightarrow}^{D_5}\mathbb{Z}^{24}
\mathop{\longrightarrow}^{D_4}\mathbb{Z}^{27}
$$
where $D_5$ and $D_4$ consist of the incidence numbers given in
Theorem \rref{TCW2}. First we derive the Smith normal form $S_4 =
U_4 D_4 V_4$ of $D_4$. It turns out that for $i=1,\ldots,13$,
${S^i}_i=1$ and all other entries vanish, so there are $24-13=11$
zero columns. Thus, $\ker D_4\cong\mathbb{Z}^{11}$ and it is
generated by the last $11$ columns of the matrix $V_4$. So the
last $11$ rows of the matrix $V_4^{-1}$ define the projection
$P^0=\left\{(V^{-1})^i_j\right\}^{i=14,\ldots,24}_{j=1,\ldots,24}$
onto $\ker D_4$. Then $D^0_5 = P^0 D_5$ is obtained by expressing
$D_5$ in the basis defined by $V$ and neglecting the first $13$
zero rows. The Smith normal form of $D_5^0$ is
$$
S_5^0= \left[ \begin{array}{cc}
\II_8 & {\bf 0}_8 \\ {\bf 0}_8^T & 3 \\ {\bf 0}_8^T & 0 \\ {\bf 0}_8^T & 0
 \end{array}\right]\,,
$$
where $\II_8$ and ${\bf 0}_8$ denote the $8$-dimensional unit
matrix and the 8-dimensional zero vector, rescpectively. Thus, we
read off $H_4(\ol\cfg_2) =
\mathbb{Z}\oplus\mathbb{Z}\oplus\mathbb{Z}_3$.

The computation of homology groups can be fully automatized by
programming the above algorithm on the computer. The program is
written in \emph{Maple 8} and uses built-in routines for computing
Smith normal forms. This way, we obtain

\btm \label{Thml}

The homology groups of $\cfg$ and of the subcomplexes $\ol\cfg_i$
are
$$
\renewcommand{\arraystretch}{1.4}
\arraycolsep5pt
 \begin{array}{c||c|c|c|c|c|c|c|c|c}
 & H_0 & H_1 & H_2 & H_3 & H_4 & H_5 & H_6 & H_7 & H_8
\\ \hline\hline
\cfg & \ZZ & 0 & 0 & 0 & 0 & 0 & 0 & 0 & \ZZ
\\ \hline
\ol\cfg_2 & \ZZ & 0 & 0 & 0 & \ZZ\oplus\ZZ\oplus\ZZ_6 & 0 & 0 & 0 & 0
\\ \hline
\ol\cfg_3 & \ZZ & 0 & \ZZ & 0 & \ZZ & 0 & 0 & 0 & 0
\\ \hline
\ol\cfg_4 & \ZZ & \ZZ\oplus\ZZ & \ZZ & 0 & 0 & 0 & 0 & 0 & 0
\\ \hline
\ol\cfg_5 & \ZZ^9 & 0 & 0 & 0 & 0 & 0 & 0 & 0 & 0
 \end{array}
$$
In particular, $\cfg$ has the homology of an $8$-sphere.
 \qed

\etm

We comment on the last statement of the theorem at the end of this section.

Next, we are going to compute the homology and cohomology groups of the
strata. We will use the following facts. Let $X$
be a compact space and let $A\subseteq X$ be a closed subset such
that $X\setminus A$ is an orientable $n$-manifold. Then
 \beq
 \label{Ghomstrata}
H_q(X\setminus A)\cong H^{n-q}(X,A;\Z)
 \,,~~~~~~
H^q(X\setminus A,\ZZ)\cong H_{n-q}(X,A)\,,
 \eeq
where $H^k(X,A;\ZZ)$ denotes the $k$-th cohomology group of $X$
relative to $A$ with coefficient group $\ZZ$, see \cite{Fuks}. We
wish to apply \eqref{Ghomstrata} to $X = \ol\cfg_i$ and $A =
\ol\cfg_{i+1}$ (thus $X\setminus A = \cfg_i$). First, applying the above algorithm
to the cell complexes $\ol\cfg_i \setminus  \ol\cfg_{i+1}$ we compute the
relative homology groups:
 \beq
 \label{Grelhml}
\renewcommand{\arraystretch}{1.4}
\arraycolsep5pt
 \begin{array}{c||c|c|c|c|c|c|c|c|c}
 & H_0 & H_1 & H_2 & H_3 & H_4 & H_5 & H_6 & H_7 & H_8
\\ \hline\hline
(\cfg,\ol\cfg_2) & 0 & 0 & 0 & 0 & 0 & \ZZ\oplus\ZZ\oplus\ZZ_3 & 0
& 0 & \ZZ
\\ \hline
(\ol\cfg_2,\ol\cfg_3) & 0 & 0 & 0 & \ZZ & \ZZ\oplus\ZZ & \ZZ & 0 &
0 & 0
\\ \hline
(\ol\cfg_3,\ol\cfg_4) & 0 & 0 & \ZZ\oplus\ZZ\oplus\ZZ_6 & 0 & \ZZ
& 0 & 0 & 0 & 0
\\ \hline
(\ol\cfg_4,\ol\cfg_5) & 0 & \ZZ^{10} & \ZZ & 0 & 0 & 0 & 0 & 0 & 0
 \end{array}
 \eeq
Second, we have to make sure that $\cfg_i$ is orientable. We will apply the
following lemma.

\ble\label{Lorib}

Let $X$ be a cell complex of dimension $n$ and let $A$ be a
subcomplex, $A\neq X$, such that $X\setminus A$ is a connected $n$-manifold. If
$H_n(X,A)\neq 0$ then $X\setminus A$ is orientable.

\ele

{\it Proof.}~ Let $C_i$ denote the $n$-cells of $X$. Let $C =
\sum_i k_i C_i$ be a chain in $X$ which is a cycle modulo $A$ and
which generates $H_n(X,A)$. There exists $i_0$ such that
$k_{i_0}\neq 0$ and $\intC_{i_0} \cap A = \emptyset$. Choose $x\in\intC_{i_0}$.
Let $i:(X,A)\to(X,X\setminus\{x\})$ denote the natural
injection. We will show that $i_\ast C\neq 0$. Then the assertion
follows from Lemma 3 in \cite[\S IV.3.3]{Schubert}. Let $f_{i_0} :
\sigma^n\to X$ denote the characteristic map of $C_{i_0}$. Choose
a subdivision $\{\sigma^n_k\}$ of $\sigma^n$ such that the simplex
$\sigma^n_0$ contains $f^{-1}_{i_0}(x)$ in its interior. By
shrinking $\sigma^n$ to $\sigma^n_0$ and composing with
$f_{i_0}|_{\sigma^n_0}$ we obtain a singular chain $C'_0 =
(\sigma^n,f'_0)$ in $X$ whose support contains $x$ and which is
embedded homeomorphically. The remaining simplices $\sigma^n_k$ of
the subdivision yield singular chains $C'_k =
(\sigma^n_k,f_{i_0}|_{\sigma^n_k})$, $k\neq 0$. By construction,
$\sum_k C'_k$ is homologous in $X$ to $C_{i_0}$, viewed as a
singular chain. Now consider $H_n(X,X\setminus\{x\})$. As $C'_0$ is embedded
homeomorphically, $i_\ast C'_0$ is a generator of $H_n(X,X\setminus\{x\})$,
hence $i_\ast C'_0\neq 0$. Since $x\in X\setminus A$ and $X\setminus A$ is an
$n$-manifold, $H_n(X,X\setminus\{x\})$ is free. Hence, $k_{i_0} i_\ast
C'_0\neq 0$. We claim that $k_{i_0} i_\ast C'_0$ and $i_\ast C$
are homologous in $(X,X\setminus\{x\})$: the singular chain $C = \sum_{i\neq
i_0} k_i C_i + k_{i_0} C_{i_0}$ is homologous in $X$ to
$\sum_{i\neq i_0} k_i C_i + k_{i_0} \sum_{k\neq 0} C'_k + k_{i_0}
C'_0$. Since all $C_i$, $i\neq i_0$, and all $C'_k$, $k\neq 0$,
are contained in $X\setminus \{x\}$, the assertion follows. This proves the
lemma.
 \qed
~\\

We check that $X=\ol\cfg_i$ and $A=\ol\cfg_{i+1}$ obey the assumptions of the
lemma. For the assumption on the homology group this follows from Table
\eqref{Grelhml}. For the assumption on connectedness, we have

\ble\label{Lstrataconnected}

The strata $\cfg_1,\dots,\cfg_4$ are connected.

\ele

{\it Proof.}~ For $\cfg_4$, the assertion is obvious. For
$\cfg_1$, we use that it is obtained from the connected cell
complex $\cfg$ by removing the subcomplex $\ol\cfg_2$ which has
codimension $3$. The same argument applies to $\cfg_3 =
\ol\cfg_3\setminus \ol\cfg_4$, because $\ol\cfg_4$ has codimension $2$ in
$\ol\cfg_3$ and from the homology groups we read off that
$\ol\cfg_3$ is connected. For $\cfg_2 = \ol\cfg_2\setminus \ol\cfg_3$, we
cannot apply this argument, because the codimension of $\ol\cfg_3$
in $\ol\cfg_2$ is $1$. Here, we use that $\ol\cfg_2$ is made up by
the nine $5$-cells $C^{221}_{ij}$, $i,j=1,2,3$. Using the boundary
formulae in Theorem \rref{TCW2} we check
$$
C^{221}_{ij}\cap C^{221}_{i\,\,\jpm} \supseteq C^{211}_{\jmp\,\,
i}
 \,,~~~~~~
C^{221}_{ij}\cap C^{221}_{\ipm\,\, j} \supseteq C^{121}_{\imp\,\,
j}\,.
$$
Hence, starting from inside $C^{221}_{11}$ one can reach any of
the $5$-cells $C^{221}_{ij}$ on a path inside the union of the
interiors of these $5$-cells and the interiors of the $4$-cells
$C^{211}_{ij}$ and $C^{121}_{ij}$, $i,j=1,2,3$. Since the latter
do not belong to $\ol\cfg_3$, this shows that $\cfg_2$ is
connected, too.
 \qed
 \\

Now, application of \eqref{Ghomstrata} yields

\btm \label{Tcohmlstrata}

The integer cohomology groups of the strata are
$$
\renewcommand{\arraystretch}{1.4}
\arraycolsep5pt
 \begin{array}{c||c|c|c|c}
 & H^0 & H^1 & H^2 & H^3
\\ \hline\hline
\cfg_1 & \ZZ & 0 & 0 & \ZZ\oplus\ZZ\oplus\ZZ_3
\\ \hline
\cfg_2 & \ZZ & \ZZ\oplus\ZZ & \ZZ & 0
\\ \hline
\cfg_3 & \ZZ & 0 & \ZZ\oplus\ZZ\oplus\ZZ_6 & 0
\\ \hline
\cfg_4 & \ZZ & \ZZ^{10} & 0 & 0
 \end{array}
$$
All the remaining cohomology groups are trivial.
 \qed

\etm

To compute the homology groups we need the relative cohomology
groups $H^k(\ol\cfg_i,\ol\cfg_{i+1};\ZZ)$. According to the
Universal Coefficient Theorem, see e.g.\ \cite[Cor.\
7.3]{Bredon:Top},
$$
H^k(\ol\cfg_i,\ol\cfg_{i+1};\Z) \cong F_k\oplus T_{k-1},
$$
where $F_k$ and $T_k$ are the free and the torsion part,
respectively, of $H_k(\ol\cfg_i,\ol\cfg_{i+1})$. We obtain
$$
\renewcommand{\arraystretch}{1.4}
\arraycolsep5pt
 \begin{array}{c||c|c|c|c|c|c|c|c|c}
 & H^0 & H^1 & H^2 & H^3 & H^4 & H^5 & H^6 & H^7 & H^8
\\ \hline\hline
(\cfg,\ol\cfg_2) & 0 & 0 & 0 & 0 & 0 & \ZZ\oplus\ZZ & \ZZ_3 & 0 &
\ZZ
\\ \hline
(\ol\cfg_2,\ol\cfg_3) & 0 & 0 & 0 & \ZZ & \ZZ\oplus\ZZ & \ZZ & 0 &
0 & 0
\\ \hline
(\ol\cfg_3,\ol\cfg_4) & 0 & 0 & \ZZ\oplus\ZZ & \ZZ_6 & \ZZ & 0 & 0
& 0 & 0
\\ \hline
(\ol\cfg_4,\ol\cfg_5) & 0 & \ZZ^{10} & \ZZ & 0 & 0 & 0 & 0 & 0 & 0
 \end{array}
$$
Then \eqref{Ghomstrata} yields

\btm \label{Thmlstrata}

The homology groups of the strata $\cfg_1,\dots,\cfg_4$ are
$$
\renewcommand{\arraystretch}{1.4}
\arraycolsep5pt
 \begin{array}{c||c|c|c|c}
 & H_0 & H_1 & H_2 & H_3
\\ \hline\hline
\cfg_1 & \ZZ & 0 & \ZZ_3 & \ZZ\oplus\ZZ
\\ \hline
\cfg_2 & \ZZ & \ZZ\oplus\ZZ & \ZZ & 0
\\ \hline
\cfg_3 & \ZZ & \ZZ_6 & \ZZ\oplus\ZZ & 0
\\ \hline
\cfg_4 & \ZZ & \ZZ^{10} & 0 & 0
 \end{array}
$$
All other homology groups are trivial.
 \qed

\etm

Next, we list those homotopy groups of
the skeleta and the closures of the strata which follow by general
theorems from our results above. Computation of further homotopy
groups remains a future task.

\bco\label{Chtp}

~\\
{\rm (i)}~ For $n\geq 2$, the $n$-skeleton $\cfg^n$ is
$(n-1)$-connected. Moreover,
$$
\pi_2(\cfg^2) = \ZZ^{14}
 ,~
\pi_3(\cfg^3) = \ZZ^{13}
 ,~
\pi_4(\cfg^4) = \ZZ^{11}
 ,~
\pi_5(\cfg^5) = \ZZ^4
 ,~
\pi_6(\cfg^6) = \ZZ^5
 ,~
\pi_7(\cfg^7) = \ZZ\,.
$$

{\rm (ii)}~ The $1$-skeleton $\cfg^1$ and the stratum $\cfg_4$ are
homotopy-equivalent to a bouquet of ten $1$-spheres. In
particular, their fundamental group is the free group on $10$
generators and the other homotopy groups are trivial.
\medskip

{\rm (iii)}~ The principal stratum $\cfg_1$ has $\pi_1(\cfg_1) =0$
and $\pi_2(\cfg_1) = \ZZ_3$.
\medskip

{\rm (iv)}~ The lower homotopy groups of the closures of the
strata are
 \begin{eqnarray*}
\pi_k(\ol\cfg_2)
 & = &
 \begin{cases}
0 &|~ k=0,\dots,3\,, \\ \ZZ\oplus\ZZ\oplus\ZZ_3 &|~ k=4\,,
 \end{cases}
\\
\pi_k(\ol\cfg_3)
 & = &
 \begin{cases}
0 &|~ k=0,1\,, \\ \ZZ &|~ k=2\,,
 \end{cases}
\\
\pi_k(\ol\cfg_4)
 & = &
 \begin{cases}
\ZZ\oplus\ZZ &|~ k=2\,, \\ 0 & |~ \mbox{otherwise.}
 \end{cases}
\end{eqnarray*}

\eco

{\it Proof.}~ (i)~ Let $n\leq 7$. Since $\cfg$ is obtained from
the $n$-skeleton $\cfg^n$ by attaching cells of dimension $n+1$
and higher, $\pi_{k}(\cfg^n) = \pi_k(\cfg) = 0$ for all $k<n$.
This is a consequence of the Whitehead theorem, see \cite[Prop.\
VII.11.6]{Bredon:Top}. Then, for $n\geq 2$, the Hurewicz theorem
implies $\pi_n(\cfg^n) \cong H_n(\cfg^n)$. The latter is just the
kernel of the boundary map in dimension $n$, hence it is a free
Abelian group. Its rank can be determined as follows. Since
$H_n(\cfg) = 0$, the rank coincides with the rank of the image of
the boundary map in dimension $n+1$, which is given by the number
of $n+1$-cells minus the rank of the kernel of the boundary map in
dimension $n+1$. Iterating this argument, we obtain that the rank
of $H_n(\cfg^n)$ is given by the alternating sum of the numbers of
cells of dimensions $n+1$ to $7$ plus/minus $1$ for the rank of
$H_8(\cfg)$.
\medskip

(ii)~ $\cfg^1$ is a graph with $9$ vertices and $18$ edges, hence
it is homotopy-equivalent to a bouquet of $(1-9+18)$ spheres of
dimension $1$. $\cfg_4$ is a $9$-punctured $2$-torus, hence it is
homotopy-equivalent to a bouquet of $10$ spheres of dimension $1$.
\medskip

(iii)~ $\cfg_1$ is obtained from $\cfg$ by removing $\ol\cfg_2$
which consists of cells of codimension $3$. Hence $\pi_i(\cfg_1) =
\pi_i(\cfg)$ for $i\leq 1$. Then the Hurewicz theorem implies
$\pi_2(\cfg_1) = H_2(\cfg_1) = \ZZ_3$.
\medskip

(iv)~ For $\ol\cfg_4$, the result is obvious because $\ol\cfg_4
\cong \mr S^1\times \mr S^1$, see Figure \rref{2-stratum} and the
corresponding explanation in the text. For $\ol\cfg_3$ we observe
that it is obtained from the $2$-skeleton $\cfg^2$ by attaching
cells of dimension $\geq3$. Hence, by \cite[Prop.\
VII.11.6]{Bredon:Top} again, it has the same $\pi_0$ and $\pi_1$
as $\cfg^2$. Both groups vanish due to (i). Then the Hurewicz
theorem implies $\pi_2(\ol\cfg_3) \cong H_2(\ol\cfg_3) = \ZZ$. For
$\ol\cfg_2$, the argument is completely analogous.
 \qed
 \\

Let us conclude with some remarks. Besides the algebraic characterizations
derived above, we have seen that the lowest dimensional strata $\cfg_5$ and
$\cfg_4$ can be identified with standard topological spaces: $\cfg_5$ is a
discrete space consisting of nine elements. $\ol\cfg_4$ is homeomorphic to a
$2$-torus and hence $\cfg_4$ is diffeomorphic to a $9$-punctured $2$-torus. 
Thus, the question arises
whether the other strata and their closures as well as the entire
reduced configuration space $\cfg$ can be expressed in
terms of standard topological building blocks like spheres or
projective spaces, too. E.g., we have found that
$\cfg$ has the homology groups of an $8$-sphere. Since it is a cell
complex and simply connected, it is then homotopy-equivalent to an $8$-sphere.
In fact, it seems likely that $\cfg$ is in fact homeomorphic to an
$8$-sphere. However, at the moment we are not able to prove or disprove this.
Note that $\cfg$ is not a differentiable manifold, hence the
famous Poincar\'e conjecture does not apply here.
The crucial point is to check whether $\cfg$ has the so-called disjoint disks
property \cite{Daverman}.

As another example, we have found that $\ol\cfg_3$ has
the homology groups of the complex projective space $\CC
P^2$. Contrary to the case of a sphere, this does not even imply that the two
spaces are homotopy-equivalent; a counterexample is provided by the bouquet of a $2$ 
and a $4$-sphere. Nevertheless, we attempted to clarify whether the two
spaces are homeomorphic but failed until now. Let us outline the strategy we
followed. If one contracts a generator of $H_2(\ol\cfg_3)$ which is a subcomplex
homeomorphic to a $2$-sphere to a point one arrives at a simply
connected quotient cell complex which has the homology groups of the
$4$-sphere and hence is homotopy-equivalent to the latter. If one could prove
that the quotient cell complex is homeomorphic to a $4$-sphere then one
would find that
$\ol\cfg_3$ is obtained by attaching a $4$-cell to a $2$-sphere.
If further one could show that the attaching map is the Hopf map
then $\ol\cfg_3$ would be homeomorphic to $\CC P^2$, indeed. 

\comment{Note that, contrary to the case of a sphere, if a
cell complex is simply connected and has the homology of $\CC P^2$,
it is not necessarily homotopy-equivalent to the latter. A
counterexample is provided by the bouquet of a $2$ and a
$4$-sphere.}

%%%%%%%%%%%%%%%%%%%%%%%%%%%%%%%%%%%%%%%%%%%%%%%%%%%%%%%%%%%%%%%%%%%%%%%%%%%%%%%%%%%%%%%
%%%%%%%%%%%%%%%%%%%%%%%%%%%%%%%%%%%%%%%%%%%%%%%%%%%%%%%%%%%%%%%%%%%%%%%%%%%%%%%%%%%%%%%

\section*{Acknowledgements}
\setcounter{equation}{0}

%%%%%%%%%%%%%%%%%%%%%%%%%%%%%%%%%%%%%%%%%%%%%%%%%%%%%%%%%%%%%%%%%%%%%%%%%%%%%%%%%%%%%%%
%%%%%%%%%%%%%%%%%%%%%%%%%%%%%%%%%%%%%%%%%%%%%%%%%%%%%%%%%%%%%%%%%%%%%%%%%%%%%%%%%%%%%%%

The authors would like to thank C.\ Fleischhack, J. Huebschmann,
J.\ Kijowski, R.\ Matthes and H.\ Toru\'nczyk for helpful suggestions and remarks.
This research was partly supported by the
Polish Ministry of Scientific
Research and Information Technology under
grant PBZ-MIN-008/P03/2003.

%%%%%%%%%%%%%%%%%%%%%%%%%%%%%%%%%%%%%%%%%%%%%%%%%%%%%%%%%%%%%%%%%%%%%%%%%%%%%%%%%%%%%%%
%%%%%%%%%%%%%%%%%%%%%%%%%%%%%%%%%%%%%%%%%%%%%%%%%%%%%%%%%%%%%%%%%%%%%%%%%%%%%%%%%%%%%%%

\appendix

%%%%%%%%%%%%%%%%%%%%%%%%%%%%%%%%%%%%%%%%%%%%%%%%%%%%%%%%%%%%%%%%%%%%%%%%%%%%%%%%%%%%%%%
%%%%%%%%%%%%%%%%%%%%%%%%%%%%%%%%%%%%%%%%%%%%%%%%%%%%%%%%%%%%%%%%%%%%%%%%%%%%%%%%%%%%%%%

%%%%%%%%%%%%%%%%%%%%%%%%%%%%%%%%%%%%%%%%%%%%%%%%%%%%%%%%%%%%%%%%%%%%%%%%%%%%%%%%%%%%%%%
%%%%%%%%%%%%%%%%%%%%%%%%%%%%%%%%%%%%%%%%%%%%%%%%%%%%%%%%%%%%%%%%%%%%%%%%%%%%%%%%%%%%%%%

\section{Table of number of cells} \label{Snr}

%%%%%%%%%%%%%%%%%%%%%%%%%%%%%%%%%%%%%%%%%%%%%%%%%%%%%%%%%%%%%%%%%%%%%%%%%%%%%%%%%%%%%%%
%%%%%%%%%%%%%%%%%%%%%%%%%%%%%%%%%%%%%%%%%%%%%%%%%%%%%%%%%%%%%%%%%%%%%%%%%%%%%%%%%%%%%%%

$$
\renewcommand{\arraystretch}{1.4}
 \begin{array}{c||c|c|c|c|c}
\dim ~~& ~~\cfg_5~~ & ~~\ol\cfg_4~~ & ~~\ol\cfg_3~~ &
~~\ol\cfg_2~~ & ~~~\cfg~~
\\ \hline\hline
0 & 9 & 9 & 9 & 9 & 9
\\ \hline
1 & - & 18 & 18 & 18 & 18
\\ \hline
2 & - & 9 & 24 & 24 & 24
\\ \hline
3 & - & - & 18 & 27 & 27
\\ \hline
4 & - & - & 6 & 24 & 24
\\ \hline
5 & - & - & - & 9 & 15
\\ \hline
6 & - & - & - & - & 9
\\ \hline
7 & - & - & - & - & 6
\\ \hline
8 & - & - & - & - & 2
\\ \hline\hline
\mbox{total} & 9 & 36 & 75 & 111 & 134
 \end{array}
$$

%%%%%%%%%%%%%%%%%%%%%%%%%%%%%%%%%%%%%%%%%%%%%%%%%%%%%%%%%%%%%%%%%%%%%%%%%%%%%%%%%%%%%%%
%%%%%%%%%%%%%%%%%%%%%%%%%%%%%%%%%%%%%%%%%%%%%%%%%%%%%%%%%%%%%%%%%%%%%%%%%%%%%%%%%%%%%%%

\section{Relation between the cells of $\TUT$ and Schubert cells of $\mr
U(3)/T$} 
\label{S-Bruhat}
\setcounter{equation}{0}

%%%%%%%%%%%%%%%%%%%%%%%%%%%%%%%%%%%%%%%%%%%%%%%%%%%%%%%%%%%%%%%%%%%%%%%%%%%%%%%%%%%%%%%
%%%%%%%%%%%%%%%%%%%%%%%%%%%%%%%%%%%%%%%%%%%%%%%%%%%%%%%%%%%%%%%%%%%%%%%%%%%%%%%%%%%%%%%

We relate the cells of $\TUT$ constructed above with the
Bruhat cells in the complexification $\GL(3,\CC)$ of $\mr U(3)$ and 
the Schubert cells in the flag manifold $\mr U(3)/T$, respectively.
Let $B \subseteq \GL(3,\CC)$ denote the subgroup of upper
triangular matrices and define $B_\tau := \tau B \tau^{-1}$,
$\tau\in S_3$. The subgroups $B_\tau$ are Borel subgroups of
$\GL(3,\CC)$ associated with the Cartan subalgebra of $\gl(3,\CC)$
of diagonal matrices and a certain choice of base for the
corresponding root system. The Bruhat cells of $\GL(3,\CC)$
relative to $B_\tau$ are the subsets $B_\tau \tau' B_\tau$,
$\tau'\in S_3$. These subsets provide a disjoint decomposition
 \beq
 \label{GBruhat}
\GL(3,\CC) = \bigcup_{\tau'\in S_3} B_\tau \tau' B_\tau\,,
 \eeq
the Bruhat decomposition. By intersection with the maximal compact
subgroup $\U(3)$, the Bruhat decomposition induces a decomposition
of $\mr U(3)$:
$$
\U(3) = \bigcup_{\tau'\in S_3} (B_\tau \tau' B_\tau)\cap \U(3)\,.
$$
Explicitly, the intersections are given by
$$
\arraycolsep10pt
\renewcommand{\arraystretch}{1.4}
 \begin{array}{r||c|c|c|c|c|c}
 & \multicolumn{6}{c}{\tau'}
\\
\tau & \tau_0 & \tau_1 & \tau_3 &
\tau_4 & \tau_5 & \tau_2
\\ \hline\hline
\tau_0 & V^0_\tau & V^1_{11} & V^1_{33} & V^2_{13} \setminus V^1_{11} \setminus
V^1_{33} & V^2_{31} \setminus V^1_{11} \setminus V^1_{33} & \U(3) \setminus
V^2_{13} \setminus V^2_{31}
\\
\tau_1 & V^0_\tau & V^1_{11} & V^1_{22} & V^2_{12} \setminus V^1_{11} \setminus
V^1_{22} & V^2_{21} \setminus V^1_{11} \setminus V^1_{22} & \U(3) \setminus
V^2_{12} \setminus V^2_{21}
\\
\tau_2 & V^0_\tau & V^1_{33} & V^1_{11} & V^2_{31} \setminus V^1_{11} \setminus
V^1_{33} & V^2_{13} \setminus V^1_{11} \setminus V^1_{33} & \U(3) \setminus
V^2_{13} \setminus V^2_{31}
\\
\tau_3 & V^0_\tau & V^1_{22} & V^1_{33} & V^2_{23} \setminus V^1_{22} \setminus
V^1_{33} & V^2_{32} \setminus V^1_{22} \setminus V^1_{33} & \U(3) \setminus
V^2_{23} \setminus V^2_{32}
\\
\tau_4 & V^0_\tau & V^1_{33} & V^1_{22} & V^2_{32} \setminus V^1_{22} \setminus
V^1_{33} & V^2_{23} \setminus V^1_{22} \setminus V^1_{33} & \U(3) \setminus
V^2_{23} \setminus V^2_{32}
\\
\tau_5 & V^0_\tau & V^1_{22} & V^1_{11} & V^2_{21} \setminus V^1_{11} \setminus
V^1_{22} & V^2_{12} \setminus V^1_{11} \setminus V^1_{22} & \U(3) \setminus
V^2_{12} \setminus V^2_{21}
 \end{array}
$$
The Bruhat cells project to open cells of the flag manifold $\mr
U(3)/T \cong \GL(3,\CC)/B_\tau$, the Schubert cells associated
with $B_\tau$. The dimensions of the Schubert cells are, in the
order of the cells as in the table, $0,2,2,4,4,6$. According to the table, when
further factorizing by left multiplication by $T$, the $0$-dimensional
Schubert cell projects to $K^0_{\tau_0}$, the $2$-dimensional
Schubert cells project to two of the $1$-dimensional cells
$K^1_{11}$, $K^1_{22}$ or $K^1_{33}$, the $4$-dimensional Schubert
cells project to two of the $2$-dimensional cells $K^2_{ij}$,
$i\neq j$, and the $6$-dimensional Bruhat cell projects to the
$4$-dimensional complement in $\TUT$ of the two $2$-cells.
Inspection of the boundaries of the two $2$-cells in $\TUT$ yields
that this complement is contractible. Although it is not a cell,
it can still be interpreted as being some $4$-disk attached to the
two $2$-cells under consideration. This way, the Bruhat cells
w.r.t.\ an arbitrary but fixed Borel subgroup $B_\tau$ of
$\GL(3,\CC)$ yield a certain cell decomposition of $\TUT$ (with
non-canonical $4$-cell though). Now consider the Bruhat
decompositions w.r.t.\ all the Borel subgroups $B_\tau$, $\tau\in
\mc S_3$. A common subdivision is given by the subsets
$$
B_\tau \sigma B_\rho
 \,,~~~~~~
\tau,\sigma,\rho\in\mc S_3
$$
(note that the labelling by three permutations contains
redundancies), see \cite{Gelfand}. Since $B_\tau \sigma B_\rho =
\tau B \sigma' B \rho^{-1}$ with $\sigma' = \tau^{-1}\sigma\rho$,
by \eqref{G-Vtau}, on passing to $\TUT$, these subsets yield any
$0$, $1$ and $2$-cell, as well as the complement of all these
cells in $\TUT$. Hence, it is this common subdivision to which the
cell decomposition constructed above corresponds on the level of
Bruhat cells.

\end{document}